%% file: main.tex
\documentclass[aps,prx,superscriptaddress,twocolumn,notitlepage,10pt,floatfix]{revtex4-2}

\usepackage{amsmath,amssymb,amsfonts}
\usepackage{mathtools}
\usepackage{physics}
\usepackage{braket}
\usepackage{bbm}
\usepackage{graphicx}
\usepackage{xcolor}
\usepackage{tikz}
\usepackage{pgfplots}
\pgfplotsset{compat=1.18}
\usepackage{booktabs}
\usepackage{array}
\usepackage{algorithm}
\usepackage{algorithmic}
\usepackage[caption=false,font=normalsize,labelfont=sf,textfont=sf]{subfig}
\usepackage{url}
\usepackage{verbatim}
\usepackage{textcomp}
\usepackage{microtype}
\usepackage{soul}
\usepackage[T1]{fontenc}
\usepackage{hyperref}

\usepackage[margin=0.75in]{geometry}
\usepackage{xcolor}
\definecolor{corec}{RGB}{60,90,160}
\definecolor{aggc}{RGB}{70,140,120}
\definecolor{edgec}{RGB}{180,120,60}
\definecolor{hostc}{RGB}{120,120,120}
\definecolor{podbox}{RGB}{235,235,235}

\tikzset{
  core/.style={draw, rounded corners, thick, fill=corec!15, minimum width=14mm, minimum height=7mm, font=\small},
  agg/.style={draw, rounded corners, thick, fill=aggc!15, minimum width=14mm, minimum height=7mm, font=\small},
  edge/.style={draw, rounded corners, thick, fill=edgec!15, minimum width=14mm, minimum height=7mm, font=\small},
  host/.style={draw, circle, thick, fill=hostc!10, minimum size=4.5mm, font=\scriptsize},
  link/.style={draw, thick},
  podframe/.style={draw, rounded corners, thick, fill=podbox, inner sep=6pt}
}
\usepackage[T1]{fontenc}
\usepackage{microtype}

\usepackage{xcolor}
\definecolor{NitishBlue}{RGB}{200,10,10}

\def\BibTeX{{\rm B\kern-.05em{\sc i\kern-.025em b}\kern-.08em
    T\kern-.1667em\lower.7ex\hbox{E}\kern-.125emX}}
\usepackage{braket}
\usepackage{hyperref}
\usepackage{physics}
\usepackage{graphicx}

\begin{document}

\title{Impact of Network Constraints on Fault-Tolerant Distributed Quantum Computing}
\author{Eneet Kaur}
\email{ekaur@cisco.com}
\affiliation{Cisco Quantum Labs, Santa Monica, CA 90404, USA}
\author{Shahrooz Pouryousef}
\affiliation{Cisco Quantum Labs, Santa Monica, CA 90404, USA}
\affiliation{Department of Computer Science and Engineering, Chalmers University of Technology, Gothenburg, Sweden}
\author{Nitish Kumar Chandra}
\affiliation{Cisco Quantum Labs, Santa Monica, CA 90404, USA}
\affiliation{Department of Informatics \& Networked Systems, School of Computing \& Information, University of Pittsburgh, Pittsburgh, PA 15260, USA}

\author{Hassan Shapourian}
\affiliation{Cisco Quantum Labs, Santa Monica, CA 90404, USA}
\author{Jiapeng Zhao}
\affiliation{Cisco Quantum Labs, Santa Monica, CA 90404, USA}
\author{Ramana Kompella}
\affiliation{Cisco Quantum Labs, Santa Monica, CA 90404, USA}
\author{Reza Nejabati}
\affiliation{Cisco Quantum Labs, Santa Monica, CA 90404, USA}
\begin{abstract}
As we move towards scalable and modular quantum computing, quantum data centres become imperative. Existing analyses typically treat network constraints in isolation or through simplified models, leaving the interplay between error correction operations and communication resources underexplored. In this work, we present an end-to-end simulation framework that jointly models surface-code operations, internal QPU connectivity, and realistic network constraints including finite entanglement generation rates, limited communication qubits, and bandwidth contention, producing execution latency, from which logical error rate estimates are obtained. The framework is modular by design, allowing individual components such as routing heuristics, scheduling policies, and network topologies to be independently replaced. Numerical evaluation reveals distinct operating regimes in which the optimal resource allocation and code distance selection shift depending on the network characteristics. These results point to tradeoffs in the design of distributed quantum computing architectures that are not visible when computation and communication are modeled separately.
\end{abstract}

\maketitle

\input{Introduction}

\input{Background}

\input{Syst}

\input{Latency_Modeling_Workflow}

\input{Numerical_Experiments}

\input{noise_metric}

\section{Conclusion}
In this work, we have presented a network-aware compilation and simulation framework for distributed surface-code. The framework takes as input a logical 
circuit, QPU connectivity, and network characteristics, and produces end-to-end 
execution latency and logical error estimates through a modular pipeline. The framework explicitly 
models finite communication qubits, entanglement generation latency, and 
network bandwidth contention, while supporting both all-to-all and grid-based internal 
QPU connectivity. An asymmetric circuit-level noise model further enables extraction of 
fault-tolerance thresholds and logical error rate estimation under distributed conditions.  By jointly modeling computation and communication constraints within a single pipeline, the framework enables exploration of architectural tradeoffs, demonstrating how ancilla allocation, communication qubit provisioning, internal connectivity, entanglement generation rates, and code distance jointly shape execution latency and logical error rates. Our results suggest that although the finite entanglement generation rates and the network-induced idle time add noise, it does not push the system out of fault tolerance regime for the considered parameters. As the quantum networks hardware and protocols improve, distributed architectures could become an attractive alternative to scaling monolithic chips alone.

Several directions remain for future work. The current framework assumes deterministic 
entanglement delivery; extending it to probabilistic protocols with per-attempt success 
probabilities, attempt-dependent fidelity, and memory decoherence during wait 
times would more faithfully capture the noise introduced during the distributed surface codes execution. Magic-state 
distillation and delivery, including factory placement and contention from 
factory-to-QPU entanglement sharing the network fabric, represent another natural 
extension. Finally, improved compilation strategies including 
joint optimization of qubit placement, routing, and network scheduling could further 
reduce distributed execution overhead beyond the heuristic approaches employed here.
\bibliography{corrected_references}

\label{sec:conclusion}
\input{appendix_section_compilation}

\input{appendix_custom}

\end{document}

%% file: Introduction.tex
\section{Introduction}

As quantum hardware matures beyond the NISQ era, the question of how to scale quantum computers to practically useful sizes has become a central challenge. Quantum algorithms offer the potential for substantial advantages over classical methods in cryptography, optimization, and quantum simulation \cite{Preskill2018,Dalzell2020,Bharti2022,Barral2025}. Recent resource estimation studies consistently indicate that fault-tolerant implementations of practically relevant algorithms require substantial number of physical qubits driven by the overhead of error correction including lattice surgery, and magic-state distillation \cite{huggins2025flasq,Leblond2023,Litinski2019,beverland2022assessing,Otten2023,vanDam2023}.

Current state-of-the-art monolithic processors, in which all qubits reside within a single device interconnected, host on the order of $10^2$–$10^3$ physical qubits~\cite{guinn2023codesigned,Malinowski2023,AbuGhanem2025, eickbusch2025demonstration, manetsch2025tweezer, dasu2026computing}. While this monolithic approach has enabled important milestones, it faces inherent scalability limitations arising from control complexity, cross-talk, fabrication yield constraints, and physical system size~\cite{manetsch2025tweezer, Stephenson2020}. These challenges have motivated a shift toward modular, networked quantum architectures \cite{VanMeter2016,Yang2023,pouryousef2026benchmarking}, in which computation is distributed across multiple quantum processing units (QPUs) interconnected by quantum and classical communication links. This modular paradigm enables large-scale scaling, providing natural fault isolation~\cite{Parnas1972,Saltzer1984,AlFares2008}.
 That is, a failure in one module does not compromise the entire system and allows faulty modules to be independently replaced without disrupting the remaining infrastructure, making it a promising path toward large-scale fault-tolerant quantum computation \cite{Wehner2018,Cuomo2020,ibm_quantum_roadmap,ionq_quantum_networks}.

Quantum computing hardware spans several distinct physical platforms, each with characteristic intra- and inter-module connectivity properties. Trapped-ion systems natively support all-to-all intra-module connectivity through the collective vibrational mode. Inter-module entanglement can be established either via photonic interfaces where ion-emitted photons are interfered to generate remote Bell pairs \cite{Monroe2014, main2025distributed}, or an ion shuttling architecture via physically moving ions across trap chains \cite{kaushal2020shuttling}. Superconducting platforms feature fixed nearest-neighbor intra-module connectivity, while inter-module scaling relies on direct microwave quantum links between chips \cite{magnard2020microwave,Song2025} or microwave-to-optical transducers for longer-range connections \cite{warner2025coherent,zhou20261km}. Neutral atom arrays offer dynamically reconfigurable intra-module connectivity through optical tweezers that physically rearrange atoms into Rydberg interaction range \cite{Bluvstein2022,Sunami2025}, with inter-module strategies based on photonic links via spin-photon entanglements \cite{van2022entangling}. The inter-module entanglement generation is typically limited.

A key implication of these limited inter-module entanglement rates and platform-dependent connectivity constraints is that quantum error correction must extend to a distributed setting. In monolithic architectures, error correction codes (e.g., surface codes) rely on fast, local interactions between neighboring qubits, enabling efficient syndrome extraction, decoding, and feedback within a tightly controlled system \cite{Gottesman1998,Fowler2012,Dennis2002}. In contrast, distributed quantum computing requires logical qubits and their associated error correction operations to span multiple, physically separated processors. This leads to the concept of distributed quantum error correction (DQEC) \cite{stack2025assessing,Jacinto2026}, where stabilizer measurements, entanglement generation, and feedforward operations must be coordinated across a network \cite{Horsman2012,Nickerson2013,Nickerson2014,Li2016,Ramette2024,Mrton2025}. Previous works have considered distributed implementation of different QEC codes including surface codes, color codes, Floquet codes and Bivariate-Bicycle codes~\cite{Ramette2024,Sutcliffe2025,Chandra2025,chandra2026distributedquantumerrorcorrection,stack2025assessing}.

 Compared to monolithic implementations, distributed error correction introduces several additional challenges. Non-local stabilizer measurements require the generation and consumption of high-fidelity entanglement between nodes \cite{Ramette2022,Ramette2024}, often under realistic conditions of loss, noise, and latency. Timing and synchronization also become more demanding, as error correction cycles must be coordinated across physically separated systems with strict temporal requirements. In addition, classical communication delays can impact real-time decoding and feedback, potentially affecting fault-tolerance thresholds. Finally, resource overheads increase due to the need for entanglement distillation, routing, buffering, and orchestration across quantum hardware \cite{DQC_cisco,pouryousef2026benchmarking}. Addressing these challenges is essential for enabling scalable, fault-tolerant distributed quantum computing systems.

Surface codes are among the most extensively studied quantum error-correcting codes, owing to their high error thresholds, compatibility with two-dimensional qubit layouts, and well-developed fault-tolerant logical operation frameworks \cite{Gottesman1998,Aharonov2006}. Extending surface-code-based fault-tolerant quantum computing to distributed architectures introduces additional architectural considerations. Depending on the design, a logical qubit may reside entirely within a single module or be distributed across multiple modules; in either case, inter-module entanglement is required  for joint stabilizer measurements in the split-qubit setting, and for entanglement-assisted logical operations such as lattice surgery \cite{Horsman2012,Mrton2025} or gate teleportation across module boundaries~\cite{Ramette2024}. 

Prior works have analyzed fault-tolerance thresholds for various modular surface-code constructions under different noise models and entanglement protocols \cite{Nickerson2013,Nickerson2014,Li2016,Ramette2024,Lin2024,stack2025assessing}. However, existing analyses typically treat network constraints in isolation or abstract them through simplified models \cite{Jacinto2026}. A unified methodology that co-models surface-code execution, internal QPU connectivity, and realistic interconnect constraints including finite entanglement generation rates, limited communication qubit provisioning, and network bandwidth contention, is currently lacking. This gap becomes increasingly consequential as we move toward deployments involving multiple QPUs and logical qubits, where network-induced overheads directly influence architectural decisions such as module sizing, interconnect bandwidth provisioning, and the allocation of communication versus computational resources. 

\vspace{10pt} \noindent\textbf{Contributions.}
In this work, we introduce a network-aware compilation and simulation framework for distributed surface-code architectures that bridges this gap by jointly modeling these constraints within a single end-to-end workflow. By capturing the interplay between computation resources and communication resources, the framework enables the evaluation of architectural tradeoffs such as module sizing, interconnect bandwidth provisioning, and communication qubit allocation.  The framework takes as input a logical circuit, ancilla allocation, and network characteristics, and simulates the end-to-end execution to obtain latency and error estimates (see Fig.~\ref{fig:latency_workflow}). The framework is modular by design: the topology-dependent compilation stage produces an annotated directed acyclic graph (DAG) which represents gate dependencies together with resource and topology constraints that the network-aware simulator consumes identically regardless of the internal connectivity model, and individual stages such as routing heuristics, ancilla allocation strategies, and scheduling policies are  separate modular components that can be independently replaced without modifying the rest of the pipeline. Specifically, our contributions are:

\begin{itemize}
    \item A compilation and simulation framework that compiles logical Clifford+T circuits to distributed lattice-surgery primitives and simulates their execution across multiple QPUs, capturing internal connectivity constraints (all-to-all and grid layouts), ancilla allocation, and network routing.
    \item Explicit modeling of network-level constraints, including finite communication qubits, entanglement generation rates, and fabric bandwidth contention within data-center topologies.
    \item Quantitative analysis of resource tradeoffs enabled by the framework, revealing how ancilla allocation, communication qubits, internal connectivity, and entanglement rates jointly determine distributed execution latency. 
    \item An asymmetric circuit-level noise model that captures the impact of entanglement generation latency on logical error rates. Slow entanglement generation stalls syndrome extraction rounds during distributed lattice surgery, increasing idle time on data qubits. We formalize this as an elevated idle depolarization rate and, combined with a seam noise model for inter-module boundaries, extract fault-tolerance thresholds and extrapolate logical error rates. 
\end{itemize}

The framework reveals several qualitative regimes governing distributed surface-code performance. The optimal ancilla allocation depends on the entanglement generation rate: when remote operations are costly, over-provisioning ancilla forces gates onto the network and degrades performance, while fast entanglement implies additional ancilla are beneficial. The scaling behavior of communication qubits differs between connectivity models: under all-to-all connectivity, the benefit depends strongly on the EPR rate, while under grid connectivity it is governed primarily by boundary placement constraints. Finally, increasing code distance improves per-gate error suppression but simultaneously increases the Bell pair demand per syndrome round, potentially pushing the system into a regime where network-induced decoherence offsets the stronger error correction. While the precise transition points depend on compilation and scheduling heuristics, the existence of these distinct operating regimes points to tradeoffs in the design of distributed quantum computing systems.
\onecolumngrid
\begin{figure}[H]
\centering
\makebox[\textwidth][c]{\includegraphics[width=0.85\textwidth]{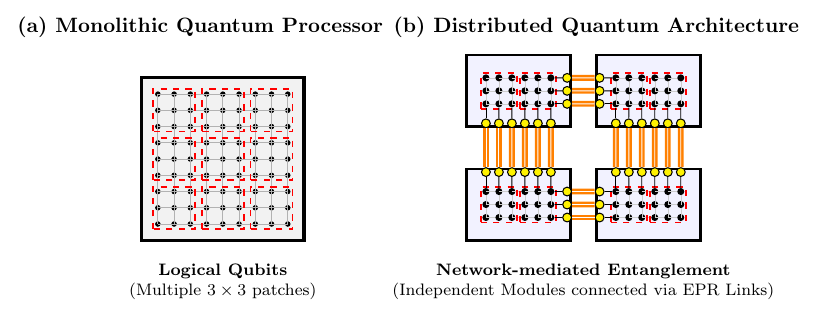}}
\makebox[\textwidth][c]{%
\begin{minipage}{0.85\textwidth}
\caption{\textbf{(a)} Monolithic single-processor design hosting all physical qubits with local nearest-neighbor connectivity. Logical qubits (dashed boundaries) are encoded within local patches. For simplicity, ancilla qubits for parity measurements are not shown. \textbf{(b)} Modular networked architecture distributing computation across multiple independent quantum processing units. Communication qubits at module boundaries enable inter-QPU entanglement distribution.}
\label{fig:architecture_comparison}
\end{minipage}}
\end{figure} 
\twocolumngrid

The paper is organized as follows. In Section~\ref{subsec:surface_codes}, we review surface codes, lattice surgery, and distributed lattice surgery. In Section~\ref{sec:architecture}, we review the quantum data-centre architecture. Section~\ref{sec:internal_qpu_connectivity} presents the internal QPU connectivity models, covering both all-to-all and grid-based layouts along with communication module placement. Section~\ref{sec:latency_model}, details the latency modeling workflow, including the compilation pipeline and the network-aware event-driven scheduler. Section~\ref{sec:experiments} presents numerical results exploring architectural tradeoffs across ancilla allocation, communication qubit provisioning, EPR generation rates, and grid sizing. Section~\ref{sec:memory_accounting} develops the asymmetric noise model and derives logical error rate estimates that incorporate network-induced latency and Bell pair infidelity. Section~\ref{sec:conclusion} concludes with a summary of findings and directions for future work.

%% file: Background.tex
\section{Surface Codes}
\label{subsec:surface_codes}

In this Section, we review surface codes, lattice surgery, and distributed lattice surgery. 
We consider fault-tolerant quantum computation based on the surface code~\cite{Fowler2012}, in which logical qubits are encoded in two-dimensional patches of physical qubits with code distance $d$. Fault-tolerant logical operations are implemented through repeated rounds of stabilizer measurements, with the number of rounds scaling as $\mathcal{O}(d)$ to reliably extract syndrome information in the presence of noisy measurements \cite{Dennis2002}. Data qubits accumulate idle errors during these rounds, which the code must also correct. For concreteness, we express all execution times in units of a single syndrome extraction round, denoted $T_{\text{syn}}$. Each syndrome round consists of ancilla initialization, a fixed sequence of physical two-qubit gates, and ancilla measurement.

In fault-tolerant quantum computation, logical Pauli operations can be efficiently handled through frame-tracking techniques \cite{Horsman2012,Fowler2012}. Under Pauli frame tracking, logical Pauli gates and phase corrections are absorbed into a classical frame via basis relabeling and post-processing, incurring no physical operations; however, Hadamard gates remain as explicit operations. The contribution of single-qubit Clifford operations to overall latency is negligible compared to two-qubit gates implemented via lattice surgery \cite{Horsman2012, Litinski2019}. The simulation framework presented here supports explicit timing models for arbitrary logical operations, including Hadamard gates, and such effects can be incorporated when required by the compilation strategy. In this work, to isolate the network impacts, we focus on quantum circuits composed of remote and local CNOT gates. We note that alternative compilation strategies such as Pauli-based computation \cite{Litinski2019} decompose circuits entirely into Pauli product rotations and Pauli product measurements, for which code surgery becomes the primary execution primitive. Our framework is compatible with such models but we focus on Pauli frame tracking framework to isolate network effects.

Non-Clifford $T$ gates are implemented via magic-state injection and gate teleportation~\cite{Bravyi2005}. Each $T$ gate consumes an injected magic state and is realized using Clifford operations, typically involving a logical CNOT between the data block and the magic state ancilla, followed by measurement and feedforward Clifford corrections. The framework can be extended to explicit modeling of magic-state delivery costs, including factory-to-QPU entanglement overhead and scenarios in which factory-to-QPU entanglement shares the network fabric with inter-QPU operations, but we defer this analysis to future work.

\subsection{Lattice Surgery} \label{subsec:lattice_surgery}
 
Fault-tolerant logical CNOT gates on surface codes can be implemented through transversal operations~\cite{Bluvstein2023} or lattice surgery~\cite{Horsman2012}. We focus exclusively on lattice surgery, which is compatible with both nearest-neighbor as well as all-to-all connectivity architectures.

Surface code patches have two boundary types, usually called $X$ and $Z$ boundaries. These boundaries define the support of the logical $X$ and $Z$ operators of the patch and provide the interfaces used in lattice surgery, where selected boundaries are merged or split to measure joint logical parities and implement entangling operations.  
A logical CNOT between a control and target qubit is realized by coupling their respective $d \times d$ surface-code patches through an intermediate ancilla patch, initialized in a known logical state. The protocol proceeds in two sequential joint parity measurements (see Fig.~\ref{fig:lattice_surgery}(a)):
 
\begin{itemize} \item \textbf{$ZZ$ measurement (merge/split):} The control and ancilla patches are coupled along their shared Z-type boundary for $d$ rounds of syndrome extraction each for the merge and split steps, implementing a joint $ZZ$ parity measurement. \item \textbf{$XX$ measurement (merge/split):} The ancilla and target patches are coupled along their shared X-type boundary for $d$  rounds of syndrome extraction each for the merge and split steps, implementing a joint $XX$ parity measurement. \end{itemize}
 
Each merge/split operation requires $2d$  syndrome rounds ($d$  for the merge, $d$  for the split), giving a total logical CNOT duration of $4d  \cdot T_{\text{syn}}$. The geometry and orientation of patch boundaries determine which parity measurements can be performed, imposing layout constraints on patch placement that become relevant in grid-based architectures (Section~\ref{subsec:grid_based}). Classically conditioned Pauli corrections determined by the measurement outcomes are tracked in the Pauli frame and incur no additional latency. The ancilla patch is reset after the operation and can be reused for subsequent gates.

\subsection{Distributed Lattice Surgery via Entanglement}
\label{sec:distributed_surgery}
 
In distributed quantum architectures, logical qubits participating in a lattice-surgery CNOT may reside on different QPUs. Since joint stabilizer measurements require physical interactions between qubits across code patches, such operations cannot be performed directly when patches are located on different QPUs. Instead, Bell pairs between communication qubits on the two QPUs are required to perform the syndrome measurements that enable the non-local joint parity measurements.
 
\begin{figure}[H]
\centering
\makebox[\textwidth][c]{\includegraphics[width=0.8\textwidth]{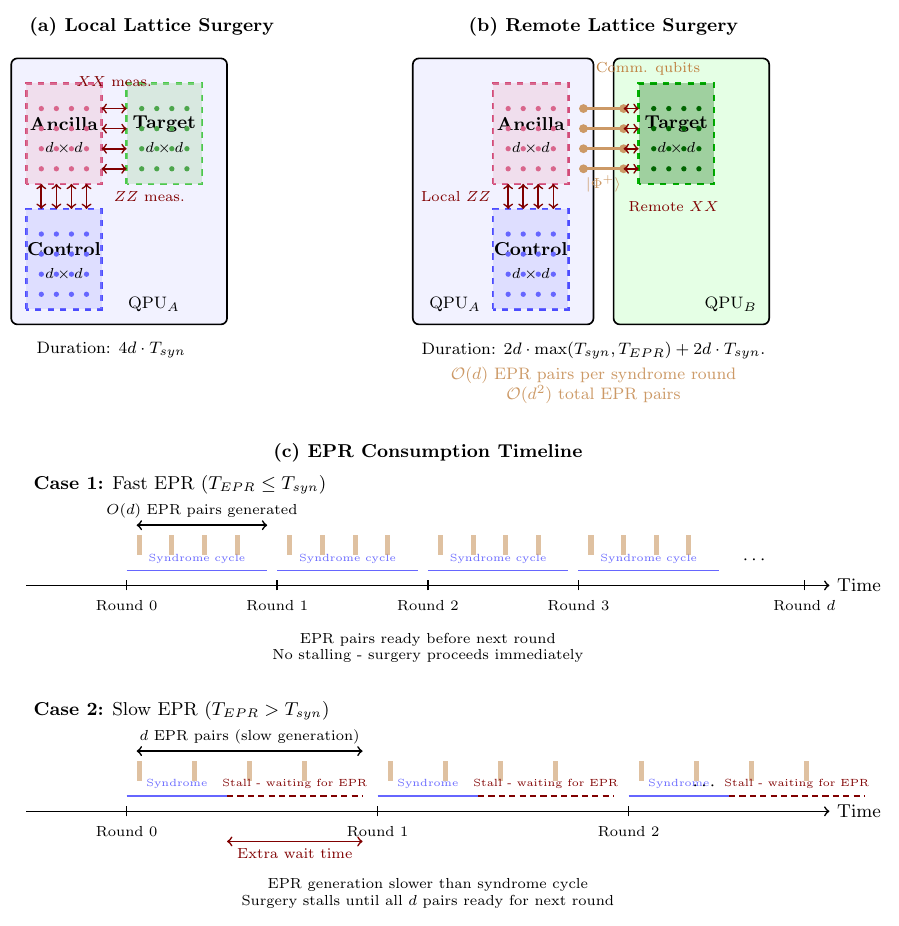}}
\makebox[\textwidth][c]{%
\begin{minipage}{0.8\textwidth}
\caption{Lattice surgery for logical CNOT operations. \textbf{(a)} Local lattice surgery: control-ancilla ($ZZ$) and ancilla-target ($XX$) measurements along patch boundaries. Duration: $4d \cdot T_{\text{syn}}$. \textbf{(b)} Remote lattice surgery across $QPU_{A}$ and $QPU_{B}$. The ancilla-target interaction requires $\mathcal{O}(d)$ Bell pairs per syndrome round via communication qubits. Total: $\mathcal{O}(d^2)$ Bell pairs. \textbf{(c)} EPR timing. Fast:$T_{\text{EPR}} \leq T_{\text{syn}}$.  Slow: $T_{\text{EPR}} > T_{\text{syn}}$: remote rounds stall, extending duration to $2d \cdot T_{\text{syn}} + 2d \cdot T_{\text{EPR}}$.}
\label{fig:lattice_surgery}
\end{minipage}}
\end{figure}

We consider the case in which the control qubit and ancilla patch reside on QPU$_A$, while the target qubit resides on QPU$_B$ (see Figure~\ref{fig:lattice_surgery}b). By construction, our implementation ensures that the ancilla patch is always co-located with either the control or the target, avoiding configurations where all three patches reside on separate QPUs. As a representative example, we consider the ancilla and control patch to be on the same QPU; the $ZZ$ measurement between control and ancilla then proceeds locally, while the $XX$ measurement between ancilla and target is mediated by entanglement.
 
Each syndrome round of the remote joint measurement requires $\mathcal{O}(d)$ Bell pairs — one for each physical CNOT gate across the patch boundary. For concreteness, we fix this count at $d$ Bell pairs per syndrome round for the remainder of this work, noting that the exact number depends on the code variant and boundary geometry. The full lattice-surgery CNOT consists of $4d$ syndrome rounds: two merge/split operations of $2d$ rounds each. Of these, $2d$ rounds correspond to the local interaction ($ZZ$ between control and ancilla) and $2d$ rounds correspond to the remote interaction ($XX$ between ancilla and target). A complete remote CNOT therefore consumes $\mathcal{O}(d^2)$ Bell pairs in total.
 
For inter-QPU communication, we assume a fixed entanglement generation time $T_{\text{EPR}}$ for producing $d$ Bell pairs between communication qubits on two QPUs. While this abstraction assumes a path-independent EPR rate, network contention is modeled explicitly: communication qubits, switch ports, and entanglement channels are finite resources, and simultaneous requests from multiple QPU pairs introduce scheduling delays. This separation isolates two distinct resource constraints: (i) the intrinsic entanglement generation rate, and (ii) network-induced delays arising from shared communication resources. Extensions to distance-dependent or topology-dependent EPR rates can be incorporated by modifying the network model without altering the workflow.

In practice, inter-module entanglement can be established through a variety of physical protocols, broadly categorized by the roles of the communicating nodes such as emitter–emitter, emitter–scatterer, or scatterer–scatterer configurations, each with different generation rates and fidelities~\cite{DQC_cisco}. These protocols may be either deterministic, producing Bell pairs on demand within a known time window, or probabilistic (attempt-until-success), where each attempt succeeds with some probability and multiple rounds may be needed before a pair is established. Regardless of the underlying protocol and physical configuration, the quantities relevant to this work can be abstracted into two parameters: the effective entanglement generation time $T_{\text{EPR}}$ (the average wall-clock time to deliver $\mathcal{O}(d)$ ready-to-use Bell pairs) and the delivered Bell pair fidelity. Throughout this work, we adopt a deterministic model in which $\mathcal{O}(d)$ Bell pairs are assumed to be available after a fixed time $T_{\text{EPR}}$, and all pairs within a syndrome round share a uniform fidelity. For probabilistic protocols, $T_{\text{EPR}}$ can be chosen large enough that the probability of successful delivery within this window approaches unity. Extending the framework to probabilistic protocols where some Bell pairs are not delivered is a natural direction for future work; this would require modeling per-pair generation latency, fidelity that depends on the number of attempts, and associated memory decoherence during wait times. Currently, these are beyond the scope of the present analysis.

The network impacts the execution of a distributed lattice-surgery CNOT in three distinct ways:
 
\begin{enumerate} \item \textbf{Resource availability delay.} Before a remote gate can begin, the required network resources such as communication qubits, switch ports, and path must be available. When multiple QPU pairs request entanglement concurrently, contention introduces waiting time before the gate can even start. The network-aware scheduler (Section~\ref{subsec:scheduling}) models this contention.

\item \textbf{Syndrome round stretching.} Once a remote gate begins, the timing of each entanglement-mediated syndrome round depends on the relationship between $T_{\text{EPR}}$ and the syndrome round duration $T_{\text{syn}} \approx 4\,T_{\text{CNOT}}$ (as syndrome extraction in surface code is performed using a depth-4 CNOT circuit, where each ancilla qubit couples to four neighboring data qubits and $T_{CNOT}$ is the time associated with CNOT gate operation.) (see Figure~\ref{fig:lattice_surgery}c):
\begin{itemize}
    \item \textbf{Fast EPR rate }($T_{\text{EPR}} \leq T_{\text{syn}}$): Bell pairs are generated within the duration of a single syndrome round. All $4d$ rounds proceed at local speed, and the remote CNOT matches the local latency:
    $$T_{\text{remote}} = 4d \cdot T_{\text{syn}}.$$
    \item \textbf{Slow EPR} ($T_{\text{EPR}} > T_{\text{syn}}$): The $2d$ local syndrome rounds proceed at local speed, but each of the $2d$ remote syndrome rounds stalls until all $d$ EPR pairs are available, stretching the effective round duration to $T_{\text{EPR}}$:
    $$T_{\text{remote}} = 2d \cdot T_{\text{syn}} + 2d \cdot T_{\text{EPR}}.$$
\end{itemize}

\item \textbf{Bell pair infidelity.} The entangled pairs used to mediate remote parity measurements are generally of lower fidelity than local two-qubit gates, introducing additional noise. This effect is captured in the asymmetric noise model (Section~\ref{sec:memory_accounting}) through an elevated error rate on seam qubits at patch boundaries.
 
\end{enumerate}
 
In the slow-EPR regime, effects $(1)$ and $(2)$ compound: scheduling delays postpone the start of the gate, and stretched syndrome rounds extend its duration, increasing idle time on all participating data qubits throughout. Additionally, the lower fidelity Bell pairs introduces additional noise at the seam qubits. We account for the combined effect of accumulated idle noise and extra seam qubits noise in our error model (Section~\ref{sec:memory_accounting}). Classical communication required for feedforward and Pauli frame updates is assumed negligible compared to entanglement generation and syndrome extraction times.
\subsection{Prior works}

Recent progress in fault-tolerant quantum computing has increasingly focused on 
modular and distributed architectures as a pathway to scalability beyond monolithic 
designs~\cite{Monroe2014, Nickerson2013}. In such architectures, logical qubits are 
encoded within individual quantum modules connected through photonic or microwave 
links, and inter-module logical operations are implemented via lattice surgery or 
entanglement-assisted protocols~\cite{Horsman2012, Mrton2025}.
Early work by Nickerson et al.~\cite{Nickerson2013, Nickerson2014} established 
fault-tolerance thresholds for surface codes distributed across modules connected 
by noisy interconnects. Li et al.~\cite{Li2016} introduced a hierarchical modular 
construction with switch-based optical connections between surface-code patches. 
Ramette et al.~\cite{Ramette2024} quantified error thresholds for entanglement-mediated 
connections at module boundaries, while Lin et al.~\cite{Lin2024} analyzed chiplet-based 
architectures with fabrication defects and demonstrated defect-aware error-correction 
strategies.

Several works have addressed partitioning and scheduling 
of quantum circuits across networked 
processors~\cite{ZomorodiMoghadam2017, AndrsMartnez2019, genetic_circ, Nikahd2021, 
Kaur2025, Baker_2020}. Ferrari et al.~\cite{Ferrari_2021} proposed a compiler for 
distributed quantum computing that employs two strategies to execute remote CNOT 
gates across quantum processors interconnected via a quantum network, and derived 
upper bounds on the compilation overhead. Cuomo et al.~\cite{Cuomo_2023} proposed 
optimized compilation strategies for general distributed programs, and 
Wu et al.~\cite{Wu_autocomm} identified burst communication patterns in distributed 
quantum programs and proposed hybrid communication schemes. However, these works 
primarily operate at the physical circuit level and do not model the interplay 
between network contention, entanglement generation rates, and fault-tolerant 
protocol timing.

At the infrastructure level, scaling quantum computing beyond single processors
has motivated the design of quantum data-center (QDC) architectures that
interconnect multiple QPUs through quantum networks~\cite{VanMeter2016}.
Several QDC topologies have been proposed, adapting classical data-center designs
such as Fat-Tree~\cite{AlFares2008}, Clos~\cite{Clos1953} and Dragonfly~\cite{Dragon_fly} to
quantum-specific constraints including stochastic entanglement generation and
contention for shared Bell-state measurement 
resources~\cite{ Sakuma2025,DQC_cisco}. Pouryousef et al.~\cite{pouryousef2026benchmarking} 
benchmarked QDC architectures (QFly, BCube, Clos, Fat-Tree) under realistic 
entanglement-generation models, characterizing latency trade-offs 
across topologies.

Despite significant progress in both distributed fault-tolerant architectures 
and quantum network modeling, a systematic framework that jointly considers 
(i)~network topology and contention, (ii)~entanglement-mediated lattice surgery 
timing, and (iii)~their combined impact on logical error rates remains lacking. 
This work bridges this gap by introducing a simulation framework that explicitly 
models the interaction between network-level resource contention and fault-tolerant 
lattice surgery execution.

\onecolumngrid
\twocolumngrid

%% file: syst.tex
\section{Network Architecture and Assumptions}
\label{sec:architecture}

In this section, we present the quantum data-center model, in which 
multiple QPUs are interconnected through a switching fabric, and detail the Fat-tree topology 
used to model network contention and bandwidth constraints.

\subsection{Quantum Data-Center Architecture}
\label{subsec:datacenter_architecture}

The quantum data centre takes inspiration from classical data centres~\cite{AlFares2008,Greenberg2009,Leiserson1985} where large-scale 
systems are organized as collections of compute nodes grouped into racks and 
interconnected through structured switching fabrics. Rather than assuming 
direct all-to-all connectivity, such architectures rely on hierarchical 
networks that trade uniform connectivity for scalability and manageability. 
A major advantage of this organization is its modular nature, providing 
resilience to single-point failures and enabling independent replacement of 
faulty components without disrupting the rest of the system.

Practical quantum architectures are similarly expected to consist of multiple 
quantum processing units (QPUs), each hosting a limited number of physical 
qubits and interfaced through a quantum interconnect, typically mediated by 
optical switches~\cite{DQC_cisco, Sakuma2025}. 
This modular approach can help overcome some of the fundamental challenges in 
scaling quantum computers, but introduces quantum communication constraints that must 
be accounted for explicitly.
\onecolumngrid
\begin{figure}[H]
\centering
\makebox[\textwidth][c]{\includegraphics[width=0.7\textwidth]{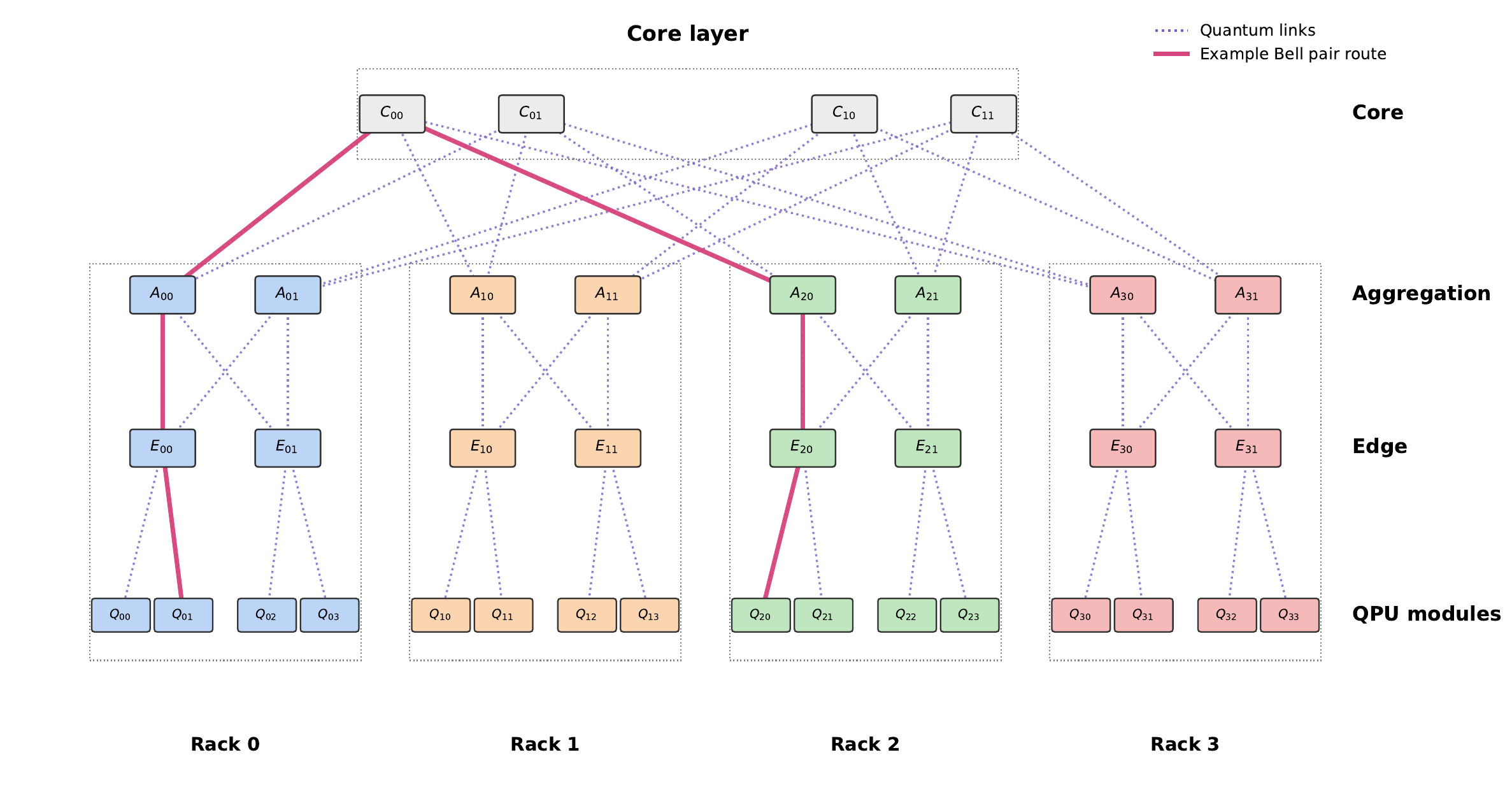}}
\makebox[\textwidth][c]{%
\begin{minipage}{0.8\textwidth}
\caption{Fat-tree topology connecting distributed quantum processing units. QPUs reside at the leaves (within pods or racks which host multiple QPUs along with edge and aggregate switches) of the network and are connected via 
    edge, aggregation, and core switches. Quantum links represent a unified physical 
    interconnect supporting both classical communication and entanglement 
    distribution.}
\label{fig:quantum-fat-tree}
\end{minipage}}
\end{figure}
\twocolumngrid
We adopt a quantum data-center model in which QPUs are organized into racks 
and interconnected through a multi-stage switching fabric. Communication 
between QPUs is mediated by a network with finite bandwidth that introduces 
both contention and delay. Each QPU is equipped with dedicated communication qubits for inter-QPU entanglement generation (Section~\ref{sec:distributed_surgery}). 
A variety of network topologies have been proposed for scalable quantum 
interconnects~\cite{Xu2025,DQC_cisco}, including Clos networks~\cite{Clos1953}, QFly 
architectures~\cite{Sakuma2025}, and Fat-tree 
topologies~\cite{AlFares2008,Alqahtani2018}. For a comparative analysis of these topologies, see Ref.~\cite{pouryousef2026benchmarking}. In this work, we focus on the Fat-tree as a 
representative data-center topology due to its widespread use, well-understood 
scaling properties, and suitability for modeling contention. The simulation framework introduced here is topology-agnostic: 
the same methodology applies to alternative network architectures by 
substituting the underlying network topology.

\subsection{Fat-Tree Interconnect}
\label{subsec:fat_tree}

A Fat-tree is a multi-rooted tree topology composed of edge, aggregation, and 
core switching layers~\cite{AlFares2008,Alqahtani2018}. QPUs reside at the leaves and 
connect to edge switches, which forward traffic through higher layers toward 
its destination. By increasing link capacity toward upper levels of the 
hierarchy, Fat-tree networks provide high bisection bandwidth and path 
diversity while maintaining scalable connectivity. 
Figure~\ref{fig:quantum-fat-tree} illustrates the topology adapted for 
distributed quantum processors.

Each network link is an optical fiber connection that carries both classical 
control traffic (scheduling, syndrome exchange, Pauli frame updates) and 
quantum channels for entanglement generation.   Physical implementations of the quantum channel may multiplex multiple channels onto a single fiber 
via wavelength or time division; however, we abstract this as a single 
effective channel per link \cite{Zhao2025}.

We assume an emitter-emitter entanglement generation protocol~\cite{Stephenson2020,Bernien2013}, in 
which each QPU's communication qubit emits a photon entangled with its local 
state. The two photons are routed to an intermediate node along the network 
path, where a Bell-state measurement (BSM) is performed. This protocol requires 
exactly one BSM per entanglement attempt. The BSM operation, together with photon transmission, introduces 
additional latency and may reduce fidelity relative to local operations. In 
our model, we absorb these effects into an effective end-to-end 
entanglement generation time $T_{\text{EPR}}$ that is assumed uniform across 
all QPU pairs. We also account for the imperfect fidelity of generated Bell pairs in Section~\ref{sec:memory_accounting}. Extensions to path-dependent or topology-dependent EPR rates, 
as well as alternative entanglement generation protocols~\cite{Awschalom2021} (e.g., 
emitter--scatterer  schemes or scatterer-scatterer), can be incorporated by modifying the network model without 
altering the compilation or scheduling workflow.

The hierarchical structure of the Fat-tree introduces contention when multiple 
QPU pairs simultaneously request entanglement along shared paths. As a result, 
the effective entanglement generation rate depends also on network topology, link bandwidth, and scheduling 
policy. This contention is modeled by the scheduler in
(Section~\ref{subsec:scheduling}).

\section{Internal QPU Connectivity Model}
\label{sec:internal_qpu_connectivity}

In this work, we model QPU internals 
exclusively at the logical level. Physical qubit layouts are abstracted away, 
and each QPU is represented as a collection of logical qubits connected by an 
effective logical connectivity graph. This abstraction allows us to isolate 
the impact of communication, routing, and scheduling on fault-tolerant computation.

For each QPU, we specify the total number of logical qubits available. A 
subset of these is designated as data qubits, while the remainder is reserved 
as ancilla qubits. Ancilla qubits are required to mediate logical two-qubit 
operations and their 
availability directly constrains the degree of achievable parallelism. We assume that the assignment of logical ancilla qubits is fixed for the 
duration of the computation, simplifying resource accounting and scheduling.

Internal logical connectivity within a QPU is represented as a graph whose 
vertices correspond to logical qubits and whose edges indicate the 
availability of two-qubit interactions between patch boundaries. Edges may 
carry additional attributes encoding constraints such as boundary 
types required for lattice-surgery operations. We specialize 
this model to two representative cases: all-to-all connectivity and 
grid-based connectivity.

\subsection{All-to-All Internal Connectivity}
\label{subsec:all_to_all}

We first consider an internal connectivity model in which logical qubits 
within a QPU are fully connected. Under this model, any logical qubit can 
participate in a local lattice-surgery CNOT with any other logical qubit 
within the same QPU, provided that an ancilla patch is available. 
Communication modules (defined in Section~\ref{subsec:comm_modules}) are 
likewise assumed to have all-to-all connectivity with the computational 
qubits within the same QPU. Geometric constraints such as boundary orientation, which arise in grid-based architectures (Section~\ref{subsec:grid_based}), are absent in this model.

This idealization is motivated by hardware platforms such as neutral atom 
arrays and trapped-ion systems, where long-range interactions or shuttling 
mechanisms can approximate all-to-all physical 
connectivity~\cite{Ramette2022,Moses2023,Bluvstein2023}. However, even on these platforms, full all-to-all 
connectivity at the logical level remains an approximation. We adopt it here 
as a deliberate modeling choice that provides an optimistic baseline, 
allowing us to isolate and study the impact of network-level 
parameters such as ancilla count, communication module count, and entanglement 
generation rate independent of internal layout constraints.

\subsection{Grid-Based Internal Connectivity}
\label{subsec:grid_based}

We next consider a grid-based internal connectivity model native to 
surface-code architectures. Logical qubits within a QPU are arranged on a 
two-dimensional grid with nearest-neighbor connectivity reflecting the 
underlying physical layout. We assume a checkerboard-style placement ~\cite{Lao2018,Akahoshi2025} of computational and ancilla logical qubits:
\[
\begin{matrix}
\text{C} & \text{A} & \text{C} & \text{A} \\
\text{A} & \text{C} & \text{A} & \text{C}
\end{matrix}
\]
where $\text{C}$ denotes a computational logical qubit and $\text{A}$ denotes 
an ancilla logical qubit. This arrangement is representative of layouts 
commonly used to support lattice-surgery-based operations. Alternative layouts optimized for specific objectives such as hexagonal routing graphs~\cite{herzog2025exploiting}, tile-based factory placement~\cite{chatterjee2025qspellbook}, and crossbar architectures~\cite{Pataki2025} have also been explored. The checkerboard arrangement adopted here provides a representative baseline for studying network-induced overheads.

Grid connectivity at the physical-qubit level induces grid connectivity at the 
logical-qubit level. As a result, logical two-qubit operations are constrained 
to nearest-neighbor interaction. Unlike the all-to-all case, the internal 
connectivity graph must explicitly track which patch boundaries are compatible 
for interaction; edges encode adjacency and also the boundary types 
that permit joint measurements required in lattice surgery. When the control 
and target qubits of a CNOT are not adjacent to a shared ancilla with 
compatible boundaries, SWAP routing through intermediate grid locations is 
required before the lattice-surgery operation can proceed.

In the grid architecture, communication modules can only be attached to boundary logical qubits due to connectivity constraints. The number of communication modules that can be attached to a given boundary patch depends on its position: corner patches, which have two exposed boundary faces, can host up to two communication modules, while non-corner boundary patches have one exposed face and can host at most one. We adopt a round-robin placement strategy that distributes communication modules across all four sides of the grid perimeter. Starting from one side, modules are placed on successive boundary patches. The modules alternate between computational and ancilla boundary patches, then placement proceeds to the adjacent side, cycling through all four sides until all allocated communication modules have been assigned or all eligible boundary positions are occupied. This ensures that communication interfaces are spread across the module rather than concentrated on a single edge, improving access for inter-QPU operations from multiple directions. Other placement strategies are possible and may yield different performance characteristics.

\subsection{Communication Modules and Entanglement Interfaces}
\label{subsec:comm_modules}

Each QPU hosts a set of dedicated \textit{communication modules} that mediate 
entanglement with other QPUs. We define a communication module as a logical 
abstraction representing the resources needed to generate $\mathcal{O}(d)$ Bell pairs per 
syndrome extraction round, where $d$ is the code distance. Physically, a 
communication module may consist of a multiple communication qubit or memory qubits that are
temporally multiplexed, a collection of communication qubits operating in 
parallel, or a spectrally multiplexed interface. Communication modules interact locally with 
computational qubits within the same QPU and remotely with communication 
modules on other QPUs that enables 
inter-QPU entanglement distribution through optical links. Examples of interface include entangled photon pairs generated by nonlinear crystals, and spin-photon entanglement with quantum frequency conversions~\cite{han2021microwave, van2022entangling, knaut2024entanglement}.\par
\input{tikz_figure}

We assume that each communication module occupies a dedicated switch port, as 
high-dimensional switches typically introduce higher loss and reconfiguration 
latency~\cite{Zhao2025}. Lattice surgery places specific constraints on 
entanglement resources: each syndrome extraction round of a joint stabilizer 
measurement requires $\mathcal{O}(d)$ Bell pairs between the participating 
remote logical qubits. By construction, each communication module provides 
exactly this capacity per round. To meet this demand within a single fiber, 
we assume that wavelength multiplexing is used to improve the entanglement 
distribution rate, significantly reducing the switch dimension requirements. 
A single remote lattice-surgery operation therefore occupies one communication 
module for its duration; multiple concurrent remote operations require multiple 
communication modules.

Entanglement generation is inherently stochastic and may be subject to 
erasures on the communication qubits when failures such as photon loss, transmission loss, detector inefficiency, or unsuccessful heralding are detected. We assume that local quantum memory times are sufficiently long that 
the stochasticity of entanglement generation does not contribute appreciably 
to overall delay, allowing us to model communication using the mean 
entanglement generation time $T_{\text{EPR}}$. While erasure events can 
affect fault tolerance and logical error rates, a detailed treatment is 
beyond the scope of this study and is deferred to future work.

%% file: tikz_figure.tex
\usetikzlibrary{arrows.meta, positioning, shapes.geometric}

\onecolumngrid
\begin{figure}[H]
\centering
\makebox[\textwidth][c]{\includegraphics[width=0.8\textwidth]{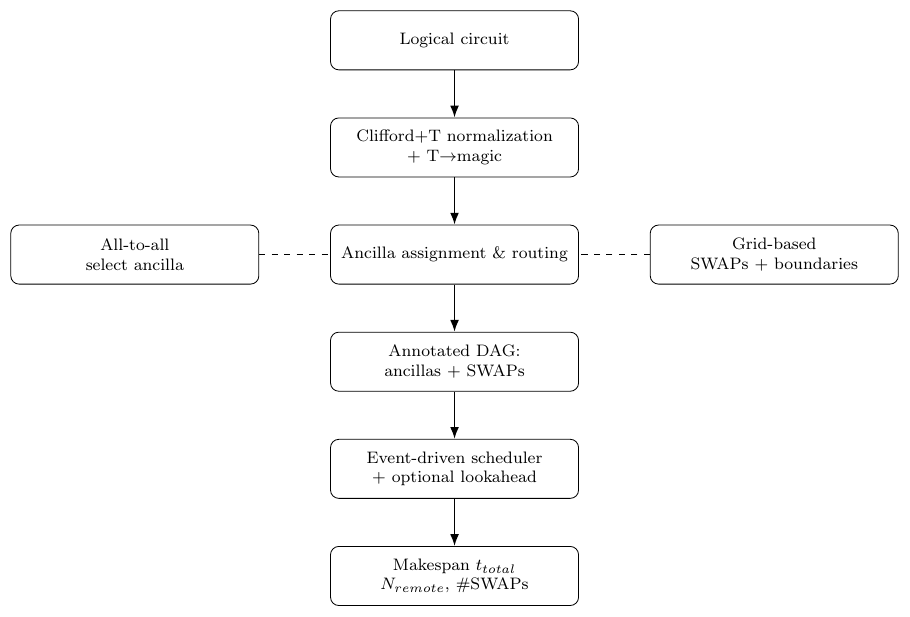}}
\makebox[\textwidth][c]{%
\begin{minipage}{0.8\textwidth}
\caption{End-to-end latency modeling workflow.}
\label{fig:latency_workflow}
\end{minipage}}
\end{figure}
\twocolumngrid

%% file: Latency_Modeling_Workflow.tex
\section{Latency Modeling Workflow}
\label{sec:latency_model}

We adopt a modular workflow for translating logical circuits into distributed 
surface-code executions. Logical operations are compiled to annotated 
lattice-surgery primitives through a topology-dependent ancilla assignment 
and routing stage, and then executed by a network-aware scheduler that 
models entanglement generation latency and resource contention. This 
separation allows us to vary architectural parameters while reusing the 
same pipeline. For the workflow, see Fig~\ref{fig:latency_workflow}.

\subsection{Compilation to Distributed Lattice-Surgery Primitives}
\label{subsec:logical_to_physical}

The compilation stage begins by taking as input a logical circuit. 
Single-qubit Clifford gates are handled as described in 
Section~\ref{subsec:surface_codes}: Pauli corrections are frame-tracked 
and incur no physical operations, while any remaining explicit single-qubit 
gates are assigned configurable execution times by the scheduler. Non-Clifford $T$ gates can be incorporated into the same workflow via 
magic-state injection and gate teleportation~\cite{Bravyi2005}: each $T$ gate 
reduces to a single lattice-surgery CNOT between the data qubit and a 
pre-distilled magic state, which the pipeline handles identically to any other 
CNOT. If the magic state resides on a remote QPU, the operation is classified 
as remote. We defer magic-state factory placement and throughput modeling to 
future work, and focus exclusively on circuits composed of logical CNOT 
operations to isolate the impact of network constraints.

The compilation of CNOT gates proceeds layer by layer over the logical 
circuit DAG. Within each layer, CNOT gates undergo an iterative ancilla 
assignment procedure: for each unassigned CNOT, the compiler identifies 
ancilla patches reachable from both the control and target data patches 
and selects the first available one. If no ancilla is available for a 
given CNOT, it is deferred. After all remaining CNOTs in that layer have been attempted, 
ancilla availability is fully reset and the deferred CNOTs are retried. 
Because previously assigned CNOTs are removed from consideration, their 
ancillas become available for reuse in subsequent iterations, modeling 
sequential execution within a single layer. This process repeats until 
all CNOTs are assigned. Between layers, all ancilla assignments are 
finalized and the process restarts.

Once an ancilla is assigned to a CNOT, the gate is expanded into its 
lattice-surgery implementation: an $M_{ZZ}$ measurement between the 
control patch and the ancilla, an $M_{XX}$ measurement between the target 
patch and the ancilla, followed by ancilla measurement and 
reinitialization. Each parity measurement is classified as local or remote 
by comparing the QPU assignments of the participating patches. The 
resulting annotated DAG, together with the timing parameters and scheduler 
configuration, forms the input to the network-aware scheduling stage 
(Section~\ref{subsec:scheduling}). Algorithm~\ref{alg:l2p} summarizes 
this procedure.

The description above assumes all-to-all internal connectivity, where 
ancilla assignment requires only selecting from the free pool. Grid-based 
architectures introduce additional constraints such as  boundary compatibility, 
routing, and communication module placement that tightly couple ancilla 
assignment to the underlying geometry. We address grid-based compilation 
separately in Appendix~\ref{appendix:grid_compilation}. In the grid-based 
case, the compilation follows the same ancilla assignment logic, but 
additionally inserts SWAP operations to route qubits to positions where 
the required lattice-surgery boundaries are available. The resulting 
annotated DAG now contains both parity measurements and SWAP gates which are then consumed by the scheduler identically to the all-to-all case.

\subsection{Network-Aware Scheduling}
\label{subsec:scheduling}
Once the logical circuit has been compiled to annotated physical primitives, 
we simulate its execution using a network-aware, event-driven scheduler. 
The scheduler extends the execution model introduced 
in~\cite{pouryousef2026benchmarking}, supporting lattice-surgery timing, 
per-qubit idle tracking, and syndrome-round stretching under slow EPR rates. 
We provide an overview here. 

The 
scheduler operates directly on the physical DAG produced by the compilation 
stage and distinguishes between operations that execute locally on a single 
QPU and operations that require inter-QPU entanglement. Each primitive is 
annotated with the QPUs involved and whether it induces a remote 
operation.

The scheduler maintains a global simulation clock and a per-qubit 
availability tracker that records when each qubit becomes free. At each 
step, the scheduler examines the frontier of the DAG and schedules each ready operation. The 
start time of an operation is determined by the latest availability time 
among its participating qubits, ensuring that no qubit is used by 
overlapping operations.

For local operations, the execution time is a fixed cost determined by the 
gate timing model. For remote operations, the scheduler additionally 
requests a network path between the source and destination QPUs, blocking 
the required resources along the path, i.e. the communication modules, switch 
ports, and shared links (Section~\ref{subsec:fat_tree}). We model a uniform entanglement generation time $T_{\text{EPR}}$ for any QPU pair. The total duration of a remote operation follows the timing model described in Section~\ref{sec:distributed_surgery}, where remote syndrome rounds can be limited by the entanglement generation time $T_{\text{EPR}}$. If the 
required network resources are not available due to contention from 
concurrent remote operations, the gate remains in the frontier and is 
retried at the next scheduling step.

Once all frontier gates have been processed, the simulation clock advances 
to the earliest completion time in the execution queue. All operations 
completing at that time are removed from the DAG and their network 
resources are released, 
exposing new frontier gates for scheduling. This process repeats until the 
DAG is empty.

The scheduler naturally captures contention for network resources: when 
multiple remote operations request entanglement concurrently, they compete 
for shared communication links, switch ports, and Bell-state measurement 
resources (Section~\ref{subsec:fat_tree}). This contention links 
architectural parameters such as communication module count, network bandwidth, 
and entanglement generation rate to observed execution latency. The 
output of the scheduling stage is the circuit makespan $t_{\text{total}}$, 
which we evaluate across a range of configurations in 
Section~\ref{sec:experiments}. Algorithm~\ref{alg:ftdqc_scheduler} summarizes 
the procedure.

%% file: Numerical_Experiments.tex
\section{Numerical Results}
\label{sec:experiments}

In this section, we evaluate the distributed surface-code execution framework 
across a range of architectural parameters. After describing the 
simulation setup, we analyze how ancilla allocation interacts with EPR 
generation rate and circuit locality to determine latency under 
all-to-all connectivity. We then examine trade offs in grid-based architectures, specifically the grid sizing vs circuit size, and finally communication module provisioning for all-to-all and grid connectivities.
\subsection{Simulation Setup}
\label{sec:sim_setup}
\begin{table}[t]
\caption{Simulation parameters.}
\label{tab:sim_params}
\scriptsize
\centering
\begin{tabular}{@{}ll@{}}
\hline\hline
\textbf{Parameter} & \textbf{Values} \\
\hline
Logical qubits ($N_q$) & $\{10,20,30,40,50\}$ \\
Gates per circuit & 200 \\
Locality ($w$) & $\{1,3,5,10,\infty\}$ \\
Circuit instances & 30 \\
\hline
QPU capacity ($Q_\text{total}$) & $\{15,20,25,30,35,40,45,50\}$ \\
Ancilla/QPU ($n_\text{ancilla}$) & $\{1,3,5,7,9,11\}$ \\
Comm./QPU ($n_\text{comm}$) & $\{1,2,4,8\}$ \\
\hline
EPR time ($T_\text{EPR}$) & $\{1,2,4,8,16\}\,T_\text{syn}$ \\
Topology & Fat-tree \\
\hline\hline
\end{tabular}
\end{table}
We evaluate the distributed surface-code execution framework described in Sections~\ref{sec:internal_qpu_connectivity} and~\ref{sec:latency_model} across a range of architectural and workload parameters.

As established in Sections~\ref{subsec:lattice_surgery} and \ref{sec:distributed_surgery}, a local CNOT has duration $T_{\text{local}} = 4d \cdot T_{\text{syn}}$, while a remote CNOT has duration $T_{\text{remote}} = 2d \cdot T_{\text{syn}} + 2d \cdot \max(T_{\text{syn}}, T_{\text{EPR}})$. Ancilla initialization and measurement can be hidden within the syndrome extraction schedule and contribute negligible additional latency.

We generate simulation circuits using a parameterized random circuit model with controlled locality. Each circuit contains 200 two-qubit gates distributed across $N_q$ logical qubits. The locality parameter $w$ controls gate placement: gates between qubits $i$ and $j$ are sampled with weight $\propto \exp(-|i - j|/w)$, where $w = 1$ produces nearest-neighbor-dominated circuits and $w = \infty$ corresponds to uniformly random qubit pairs. This parameter directly affects the fraction of gates that become remote under a given qubit-to-QPU mapping, allowing us to study how circuit locality interacts with network constraints. For each configuration, results are averaged over 30 independent circuit instances to ensure statistical reliability. Error bars in all figures represent the standard error of the mean (SEM) across these instances. The relative ordering and trends across configurations are stable as the same circuit instances are reused across parameter sweeps. 

Table~\ref{tab:sim_params} summarizes the simulation parameters. The quantum data-center architecture consists of QPUs arranged in a Fat-tree network topology (Section~\ref{subsec:fat_tree}). Each QPU has a fixed total logical qubit capacity, partitioned into computational and ancilla. The effective computational capacity of a QPU is $Q_\text{eff} = Q_\text{total} - n_\text{ancilla}$.

\subsection{Numerical analysis: all-to-all}

\subsubsection{Ancilla Allocation}
\label{sec:ancilla_allocation}

We use the simulation framework to explore how ancilla allocation interacts with 
entanglement generation latency in determining the makespan of the circuit. Increasing the ancilla allocation improves parallelism by 
enabling more concurrent lattice-surgery operations, but simultaneously 
reduces the effective computational capacity 
$Q_\text{eff} = Q_\text{total} - n_\text{ancilla}$. As $Q_\text{eff}$ 
decreases, the remote gates increases and the circuit is exposed to network-induced latency.

\begin{figure*}[t]
\centering
\includegraphics[width=\textwidth]{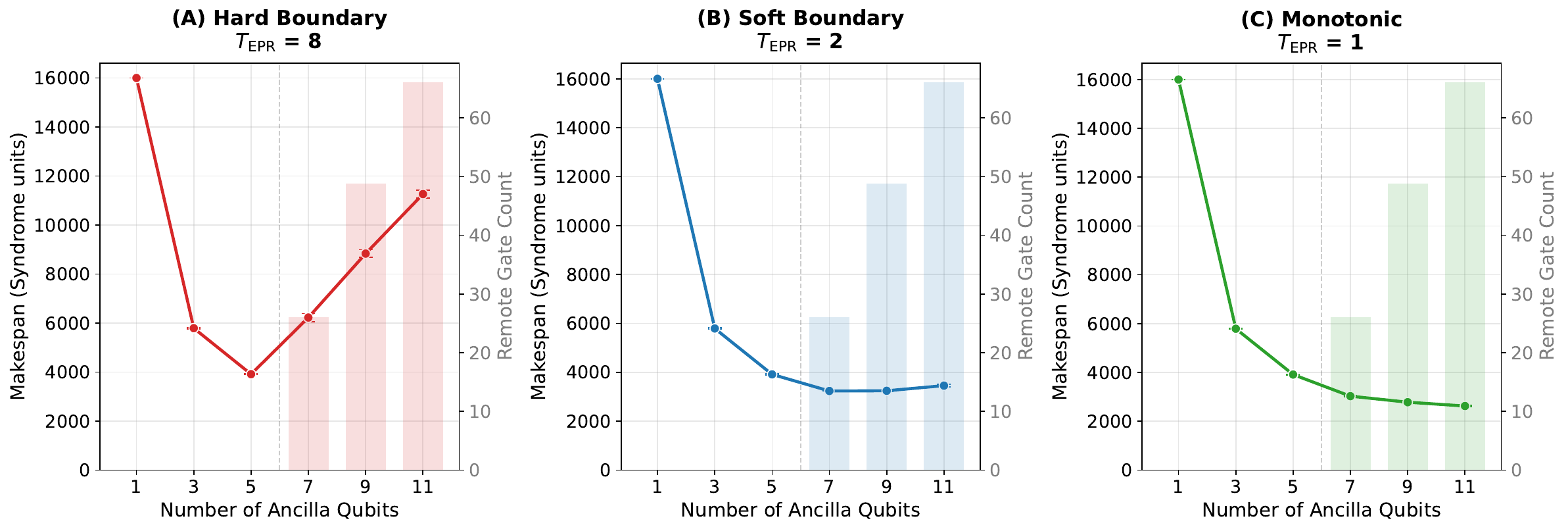}
\caption{This figure shows the ancilla scaling regimes. The parameters of the circuit are 30 logical qubits, $w= \infty$, $Q_\textrm{comp} = 35$, two communication qubits and $d=20$. }
\label{fig:phenomenon}
\end{figure*}

Figure~\ref{fig:phenomenon} illustrates three qualitatively distinct 
scaling behaviors that emerge from this tradeoff. The key variable is 
the ratio of entanglement generation time ratio, 
$T_\text{EPR} / T_\text{syn}$:

\begin{itemize}
    \item \textbf{Hard boundary} 
    ($T_\text{EPR} \gg T_\text{syn}$, panel A): The high cost of each 
    remote gate creates a sharp optimum. Increasing ancilla count 
    initially improves intra-QPU parallelism, but once $Q_\text{eff}$ 
    drops below the circuit size, the remote gate count increases and the makespan increases 
    sharply. Beyond this point, increasing ancillas is harmful.
    
    \item \textbf{Soft boundary} 
    ($T_\text{EPR} \geq T_\text{syn}$, panel B): The moderate 
    communication penalty allows the parallelism benefit to compensate 
    for remote gate overhead, 
    producing a broad optimum near the local-to-distributed transition.
    
    \item \textbf{Monotonic} 
    ($T_\text{EPR} \lesssim T_\text{syn}$, panel C): Remote gates 
    incur negligible additional cost, and increasing ancilla count is 
    monotonically beneficial across the tested range.
\end{itemize}

When the circuit is inherently distributed, for example, when the 
number of logical qubits exceeds the capacity of a single QPU even with low ancilla, the 
makespan is dominated by remote gate latency from the outset, and the 
scaling is monotonically improving \textit{regardless} of EPR rate. These results demonstrate that the optimal ancilla allocation is not fixed but varies across operating regimes. 

\subsubsection{Parameter Space Exploration}
\label{sec:param_exploration}
\begin{figure*}[h!]
\centering
\includegraphics[width=\textwidth]{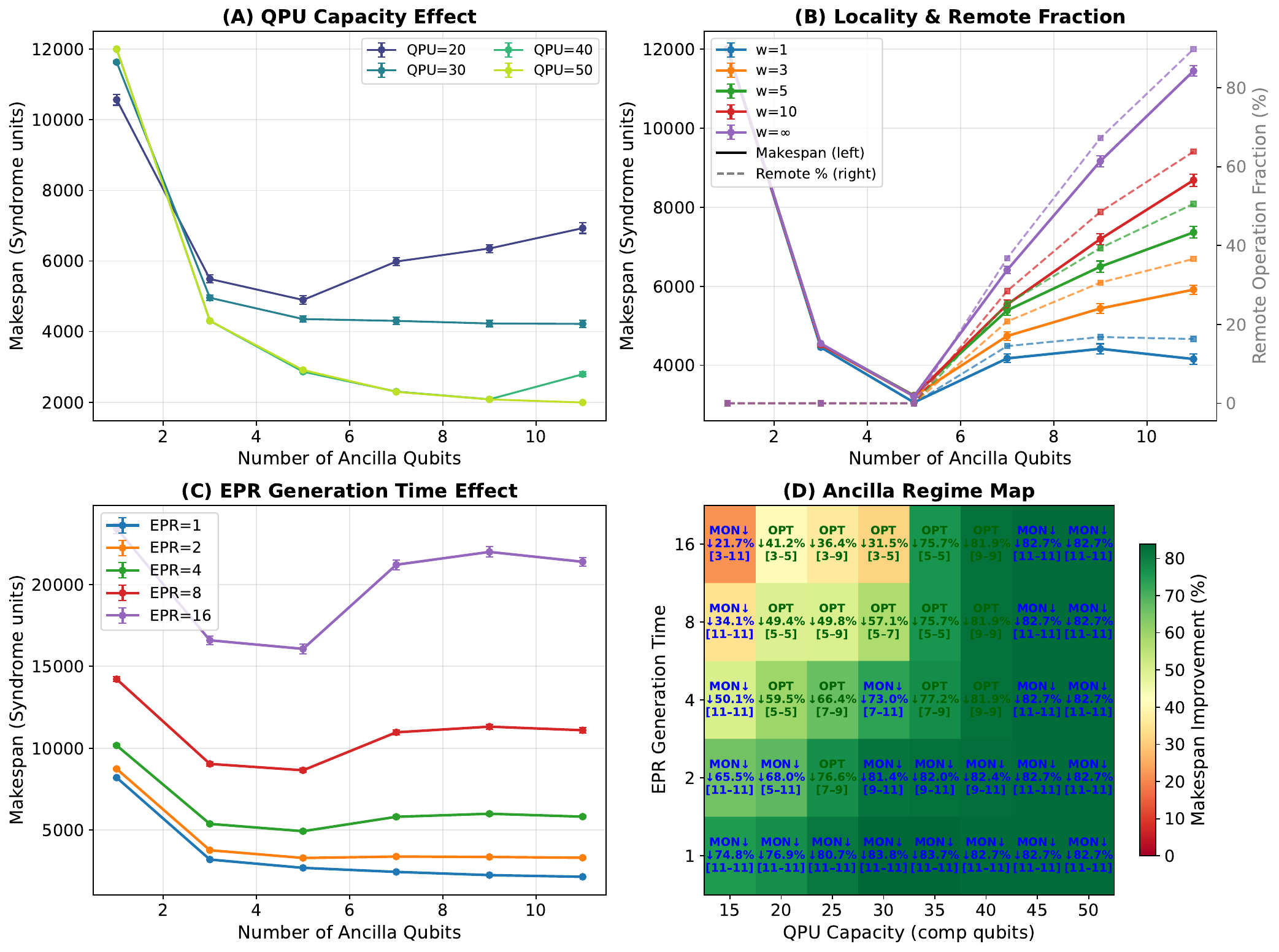}
\caption{\textbf{Ancilla scaling in distributed quantum computing.}
    All panels use circuits with 200 gates, code distance $d=15$, and 
    $n_{\mathrm{comm}} = 2$.
    \textbf{(A)}~QPU capacity effect: 30-qubit circuits with locality 
    $w=3$ and $T_{\mathrm{EPR}}=8\,T_{\mathrm{syn}}$.
    \textbf{(B)}~Locality and remote fraction: 20-qubit circuits with 
    $Q_{\mathrm{total}}=25$ and $T_{\mathrm{EPR}}=8\,T_{\mathrm{syn}}$.
    \textbf{(C)}~EPR generation time effect: 20-qubit circuits with 
    $w=5$ and $Q_{\mathrm{total}}=15$.
    \textbf{(D)}~Ancilla regime map: 30-qubit circuits with $w=5$; 
    each cell shows the regime classification (MON or OPT), percentage 
    makespan improvement over the baseline ($n_{\mathrm{ancilla}}=1$), 
    and the equivalence band 
    $[a_{\min}\text{--}a_{\max}]$ of ancilla counts statistically 
    indistinguishable from the optimum 
    ($p > 0.05$, pairwise $z$-test).}
\label{fig:param_space}
\end{figure*}

We now use the framework to systematically vary architectural parameters 
and examine their impact on distributed execution latency. 
Figure~\ref{fig:param_space} presents four complementary perspectives on 
resource tradeoffs.

Panel~A varies QPU capacity for a fixed circuit size. Smaller QPUs 
exhibit sharper optima: the transition from local to distributed 
execution occurs at lower ancilla counts, producing a more pronounced 
makespan penalty beyond the optimum. Larger QPUs display weaker 
non-monotonicity, as the circuit can be accommodated with sufficient 
margin even at higher ancilla budgets. QPU capacity 20 achieves 
slightly lower makespan than larger capacities. This occurs because a 
20-qubit QPU forces the 30-qubit circuit onto two QPUs, each contributing 
one ancilla enabling parallelism. While larger QPUs accommodate the full circuit on a single QPU with 
only one ancilla.

\text{Panel}~B overlays makespan  with the remote operation fraction as ancilla count increases across different locality settings. The correspondence 
is direct: configurations where the remote fraction rises steeply also 
exhibit rising makespan, confirming that the local-to-remote transition 
is the primary driver of the non-monotonic behavior observed in Panel~A.

Panel~C varies the entanglement generation time for a fixed circuit and QPU configuration ($N_q = 20$, $Q_\text{total} = 15$). In the slow-EPR regime ($T_\text{EPR} = 8, 16$), a clear optimum emerges near $n_\text{ancilla} = 5$, corresponding to a balanced partition of 10 computational qubits and 5 ancilla qubits per QPU. As the EPR rate improves ($T_\text{EPR} \lesssim 2$), the communication penalty diminishes and we observe that increasing ancilla has low impact on makespan.

Panel~D summarizes the scaling behavior across the full 
(QPU capacity $\times$ EPR rate) parameter space. Each cell is 
classified as MON (monotonically improving at maximum ancilla count) or 
OPT (a clear optimum exists within the tested range), along with the 
percentage makespan improvement over the single-ancilla baseline and 
the equivalence band of ancilla counts statistically indistinguishable 
from the optimum ($p > 0.05$, pairwise $z$-test). At small QPU 
capacities, the circuit is heavily distributed at all ancilla counts 
and the regime is uniformly monotonic. The OPT regime emerges at 
moderate capacities under slow-to-moderate EPR rates, where increasing 
ancilla count triggers a transition from local to distributed execution. 
At large capacities, the regime returns to monotonic, as additional 
ancillas consistently improve concurrency without forcing significant 
distribution. 
\begin{figure}
\centering
{%
  \includegraphics[width=\columnwidth]{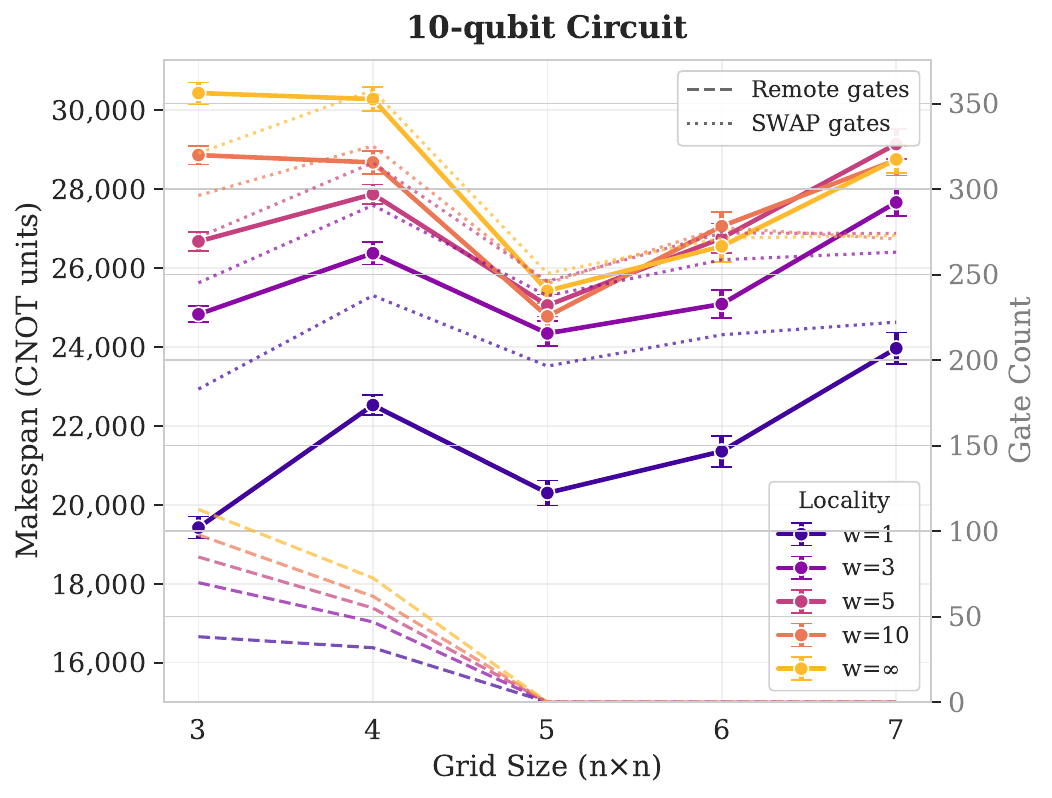}%
}\\[6pt]
{\includegraphics[width=\columnwidth]{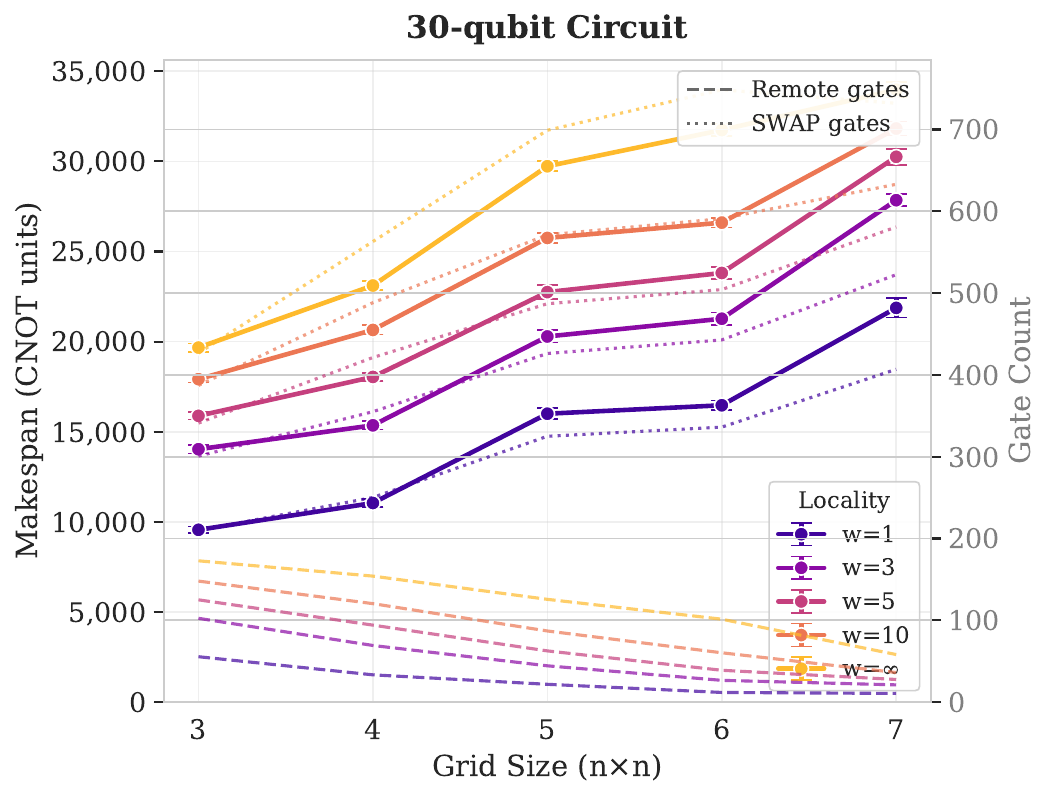}%
}
\caption{Grid size scaling at $T_\mathrm{EPR}=8$. Makespan versus grid size. For 10 qubits, remote reduction outweighs SWAP growth at intermediate sizes, producing non-monotonic behavior. For 30 qubits, SWAP growth dominates and makespan increases monotonically.}
\label{fig:grid_scaling}
\end{figure}
\subsection{Grid-Based Connectivity}

\label{subsec:grid-scaling}
We analyze grid-based internal connectivity, where logical patches are arranged on an $m \times m$ lattice within each QPU. Unlike the all-to-all connectivity model, two-qubit gates might require SWAP gates where each logical SWAP operation which decomposes into three sequential lattice-surgery 
CNOTs at a cost of $12d \cdot T_{\text{syn}}$ per SWAP operation. Increasing grid size creates a direct tradeoff: larger modules reduce the fraction of remote CNOT operations, but increase average routing distance and hence SWAP operation overhead. Both local and remote gates incur this cost, as even remote CNOTs require logical qubits to be routed to boundary patches hosting communication modules.

The relative scaling of these two competing effects determine whether the grid size increase is beneficial or detrimental. Figure~\ref{fig:grid_scaling}(a) illustrates the regime where remote gate reduction dominates: as grid size increases from $3 \times 3$ to $5 \times 5$, the remote gate count decreases substantially while SWAP growth remains modest, producing non-monotonic makespan behavior with a clear optimum at intermediate grid sizes. Figure~\ref{fig:grid_scaling}(b) illustrates the opposite regime where SWAP overhead grows more rapidly than remote gates are eliminated. Both circuits use a fixed gate count of 200. The 30-qubit circuit has lower depth due to greater parallelism, resulting in lower base makespan.

The observed scaling depends on the compilation heuristic; improved 
placement or routing strategies could reduce SWAP overhead, particularly 
for larger grids. The qualitative tradeoff between SWAP cost and remote 
gate reduction, however, is fundamental to grid-based architectures and 
persists across $T_{\text{EPR}} \in \{4, 8, 16, 32\}$.

\begin{figure}
\centering
\subfloat[Marginal improvement (all-to-all)\label{fig:comm_marginal}]{%
  \includegraphics[width=\columnwidth]{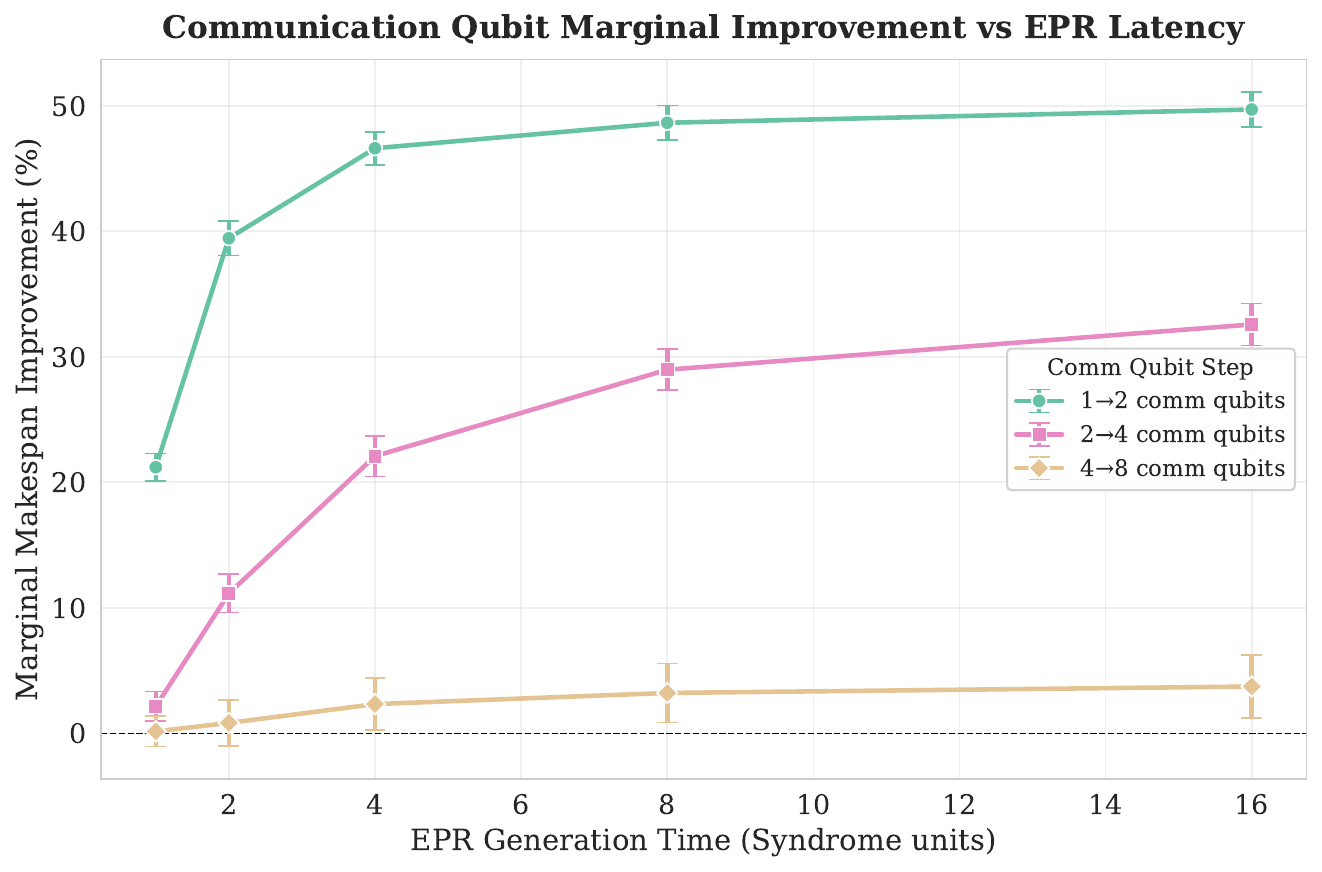}%
}\\[6pt]
\subfloat[Grid connectivity\label{fig:comm_grid}]{%
  \includegraphics[width=\columnwidth]{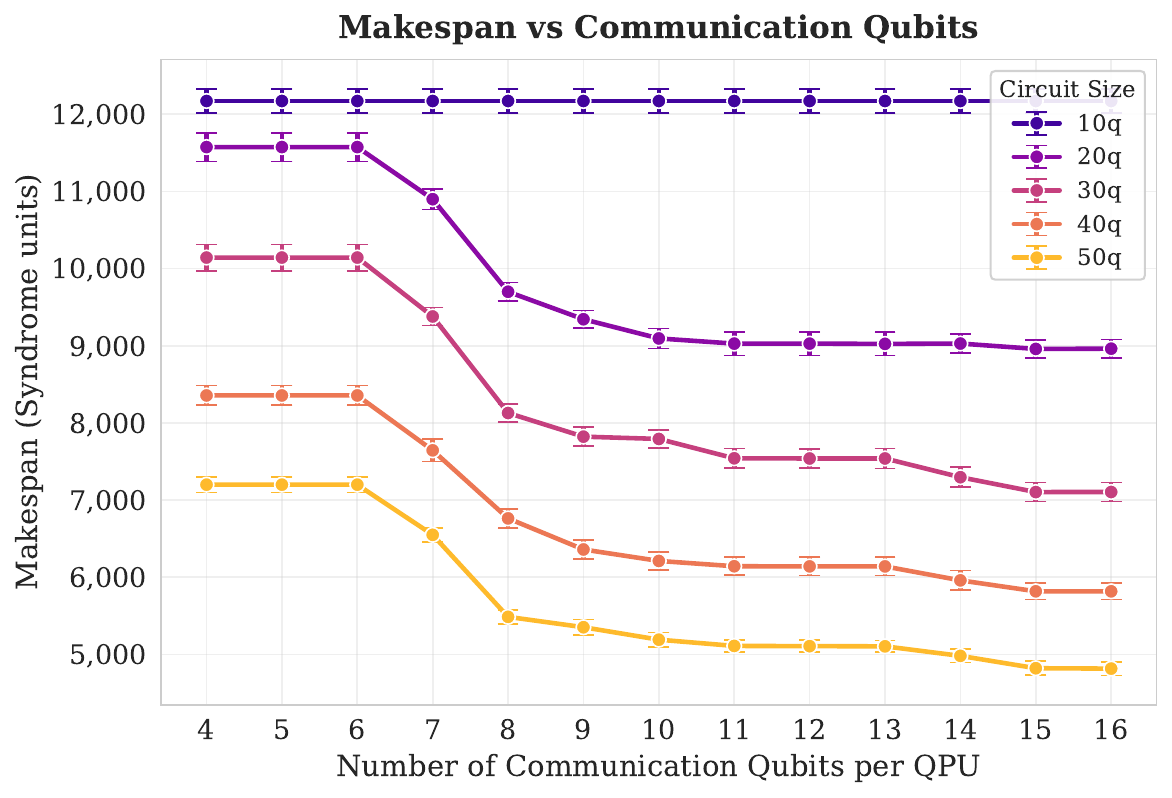}%
}
\caption{Communication qubit scaling under all-to-all and grid connectivity.
(a)~Marginal makespan improvement when doubling communication qubits per QPU ($1 \to 2$, $2 \to 4$, $4 \to 8$) as a function of EPR generation time under all-to-all connectivity. Configuration: 30 qubits, $w=5$, $Q_\text{total}=15$, $n_\text{ancilla}=5$, $d=15$, BSM capacity$=8$. 
(b)~Makespan versus communication qubits per QPU for varying circuit sizes under grid connectivity ($5 \times 5$ grid, $T_\mathrm{EPR}=8$, $w=3$).}
\label{fig:comm_scaling}
\end{figure}

\subsection{Communication Qubits Provisioning}
\label{sec:comm_provisioning}
We next examine how the number of communication qubit interacts with entanglement generation latency and internal connectivity. Figure~\ref{fig:comm_scaling} presents the tradeoffs under all-to-all and grid-based  architectures.

Under all-to-all connectivity, Figure \ref{fig:comm_scaling}, the benefit of additional communication qubits depends strongly on $T_\text{EPR}$. When $T_\text{EPR} \lesssim T_\text{syn}$, remote operations release their communication qubits quickly, limiting contention. As $T_\text{EPR}$ grows, each remote operation occupies its qubit for longer duration, stalling concurrent gates. Additional communication qubits alleviate this, with the largest gain when communication qubits grows from $1 \to 2$ and diminishing returns thereafter as the system approaches the BSM-limited regime (where the number of BSM's per switch becomes a restriction).

Under grid connectivity, the behavior differs. For small circuits the remote fraction is zero and makespan is independent of communication qubits. As circuit size grows, communication qubits become critical, but unlike the all-to-all connectivity case, the benefit is largely insensitive to $T_\text{EPR}$: each communication qubit maps to a distinct boundary patch, so saturation is set by the number of boundary locations needed to serve the circuit's remote demand rather than by entanglement generation rate.  Notably, makespan continues to improve modestly beyond the BSM per switch (8 in this configuration). This occurs because additional communication modules at distinct boundary positions reduce the average SWAP distance that a qubit must reach to access the communication qubit.

%% file: noise_metric.tex
\section{From Latency to Logical Error Rates}
\label{sec:memory_accounting}

The scheduling framework developed in Sections~\ref{sec:architecture}--\ref{sec:latency_model} produces, for each circuit execution, the makespan and per-qubit operation information such as the idle time and gate operations (both local and non-local). In this section, we describe how the scheduler output can be translated into total logical error rate estimates (TLER) via an analytic approximation.

We use an approximate ansatz because the TLER of a fault tolerant circuit  generally does not admit a simple closed form expression. Estimating it requires specifying the syndrome extraction procedure, noise model, and decoder, and tracking how physical errors propagate through the full spacetime volume of FTQC~\cite{Tan2024}. This analysis becomes computationally expensive, and often intractable, at large code distances.
We instead work with an additive ansatz that decomposes the TLER into independent per-gate and per-idle-period contributions. This approximation is suited to the fault-tolerant regime, where individual logical error rates are small, the total operation count $N$ satisfies $N \cdot \text{LER} \ll 1$, and error correction maintains approximate independence between error events.

Networks contribute to the TLER through three distinct mechanisms:
\begin{enumerate}
    \item \textbf{Extended idle times.} Scheduling delays and entanglement generation waiting times increase the total idle duration experienced by all logical qubits, requiring additional memory protection rounds and accumulating memory errors.
    \item \textbf{Stretched syndrome rounds during remote gates.} When $T_{\text{EPR}} > T_{\text{syn}}$, the entanglement-mediated syndrome rounds of a remote lattice-surgery operation are delayed, extending the effective round duration. Data qubits participating in the gate experience elevated idle depolarization during these stretched rounds. This effect is captured by the asymmetric noise model (Appendix~\ref{appendix:noise_in_lattice_surgery}) through the idle noise rescaling parameter $k$, and is reflected in the per-gate $\text{LER}_g$ rather than in the memory accounting. See Fig~\ref{fig:lattice_surgery} for the schematic on connectivity of surface code patches and the associated timing parameters.

    \item \textbf{Bell pair infidelity.} Entangled pairs used to mediate remote parity 
    measurements are generally of lower fidelity than local two-qubit gates, introducing 
    additional noise at inter-module boundaries. This effect is captured by the seam noise 
    parameter $\lambda$ in the asymmetric noise model 
    (Appendix~\ref{appendix:noise_in_lattice_surgery}) and enters the TLER through the 
    per-gate $\text{LER}_g$ term.
    
\end{enumerate}

We approximate the TLER by decomposing it into independent per-gate and 
per-idle-period contributions:
\begin{equation}
\text{TLER} \approx \sum_{g \in \text{Gates}} \text{LER}_g + \sum_{i=1}^{n} \text{LER}_{\text{memory}, i}
\label{eq:tler_proxy}
\end{equation}
where $\text{LER}_g$ is the logical error rate for gate $g$, and $\text{LER}_{\text{memory}, i}$ is the accumulated memory error for qubit~$i$ during idle periods. This additive decomposition relies on $\prod(1-x_i) \approx 1 - \sum x_i$, valid when $N \cdot \text{LER} \ll 1$.

\subsection{Memory Round Accounting}
\label{sec:memory_rounds}

To maintain fault-tolerance during idling, the syndrome extraction process should be repeated \(d\) times to ensure the temporal distance matches the code's spatial distance. This enables the decoder to protect the logical state against both measurement errors and data qubit errors between active operations.
Memory experiment therefore consists of $d$ consecutive syndrome extraction rounds, with duration $T_{\text{mem}} = d \cdot T_{\text{syn}}$, where $T_{\text{syn}}$ is the duration of a single syndrome extraction cycle. We denote by $\widetilde{\text{LER}}_{\text{mem}}$ the logical error rate associated with one complete $d$-round memory experiment.

For each logical qubit $i$, let $T_{\text{idle},i}$ denote the total time the qubit spends idle during circuit execution. The number of memory experiments required is:
\begin{equation}
N_{\text{mem}, i} = \left\lceil \frac{T_{\text{idle}, i}}{d \cdot T_{\text{syn}}} \right\rceil,
\label{eq:nmem}
\end{equation}

where the ceiling function \(\lceil x \rceil\) represents the smallest integer greater than or equal to \(x\).

 The idle time $T_{\text{idle},i}$ represents the \textit{total accumulated} idle duration 
for qubit~$i$ across the entire circuit execution, aggregated over all intervals in which 
the qubit is not participating in a gate operation. Rather than tracking each contiguous 
idle interval separately and modeling the syndrome extraction state at individual 
gate boundaries, we pool the total idle time per qubit and divide once by the memory 
experiment duration. This is an approximation: a more granular treatment would account 
for the placement and length of each idle interval individually, potentially yielding 
tighter error estimates.

Defining $N_{\text{mem}} = \sum_{i=1}^{n} N_{\text{mem},i}$ as the total number of memory experiments across all qubits, the total circuit error proxy becomes:
\begin{equation}
\text{TLER} \approx \sum_{g \in \text{Gates}} \text{LER}_g \;+\; N_{\text{mem}} \cdot \widetilde{\text{LER}}_{\text{mem}}
\label{eq:total_ler}
\end{equation}

\subsection{Per-Gate Logical Error Rates}
\label{sec:per_gate_ler}

A logical CNOT operation performed using lattice surgery consists of two sequential joint parity 
measurements $M_{ZZ}$ and $M_{XX}$ each of which 
can produce either an $X_L$ or $Z_L$ logical failure. For the rotated surface code under 
depolarizing noise, $p_{X_L} = p_{Z_L}$ by symmetry, so we characterize both channels by 
$p_{X_L}$.

For idle memory, each $d$-round memory experiment contributes one $X_L$ and one $Z_L$ 
failure channel:
\begin{equation}
    \widetilde{\text{LER}}_{\text{mem}} = 2\, p_{X_L}^{\text{local}}.
\end{equation}

For a \textit{local} CNOT, both measurements proceed under the baseline noise model (See Appendix~\ref{appendix:noise_in_lattice_surgery}). Each 
measurement contributes two failure channels ($X_L$ and $Z_L$), giving:
\begin{equation}
    \text{LER}_g^{\text{local}} = 4\, p_{X_L}^{\text{local}}.
\end{equation}

For a \textit{remote} CNOT, the $M_{ZZ}$ measurement between control and ancilla proceeds 
locally, while the $M_{XX}$ measurement between ancilla and remote target is mediated by 
entanglement across the inter-module seam. Given this orientation, only the $X_L$ component is sensitive to the seam boundary, while the $Z_L$ component is determined by bulk stabilizers unaffected by the inter-module link \cite{Ramette2024}. The $X_L$ component of this non-local 
measurement experiences both elevated idle noise ($k$) during stretched syndrome rounds 
and reduced entanglement fidelity ($\lambda$), yielding $p_{X_L}^{\text{remote}}(k,\lambda)$. 
The remaining three error channels are unaffected by the seam:
\begin{equation}
    \text{LER}_g^{\text{remote}} = 3\, p_{X_L}^{\text{local}} + p_{X_L}^{\text{remote}}(k,\lambda).
\end{equation}

Both logical error rates follow the sub-threshold scaling ansatz 
$p_{X_L} = A\,(p/p_{\text{th}})^{(d+1)/2}$, with parameters determined by the noise 
regime. For local operations and idle memory, we use baseline parameters 
$p_{\text{th}}^{\text{base}} = 0.743\%$ and $A^{\text{base}} = 0.05$ from Ref.~\cite{Jacinto2026}. For the non-local 
$X_L$ component, we use seam-aware parameters $A(k,\lambda)$ and $p_{\text{th}}(k,\lambda)$ 
extracted from the threshold sweep (Appendix~\ref{appendix:noise_in_lattice_surgery}, 
Figure~\ref{fig:threshold_sweep}). Combining all contributions:
\begin{align}
    \text{TLER} \approx &N_{\text{local}}\, 4 p_{X_L}^{\text{local}} 
    + N_{\text{remote}} \left(3 p_{X_L}^{\text{local}} + p_{X_L}^{\text{remote}}(k,\lambda)\right) 
    + \nonumber \\&N_{\text{mem}}\, 2 p_{X_L}^{\text{local}},
\label{eq:tler_full}
\end{align}
where $N_{\mathrm{local}}$ and $N_{\mathrm{remote}}$ are the local and remote
CNOT counts produced by the compilation stage (Section~\ref{subsec:logical_to_physical}),
and $T_{\mathrm{idle}}$ is the total per-qubit idle time reported by the
network-aware scheduler (Section~\ref{subsec:scheduling}).

\subsection{Numerics for TLER}
\begin{figure}
    \centering
\includegraphics[width=0.9\linewidth]{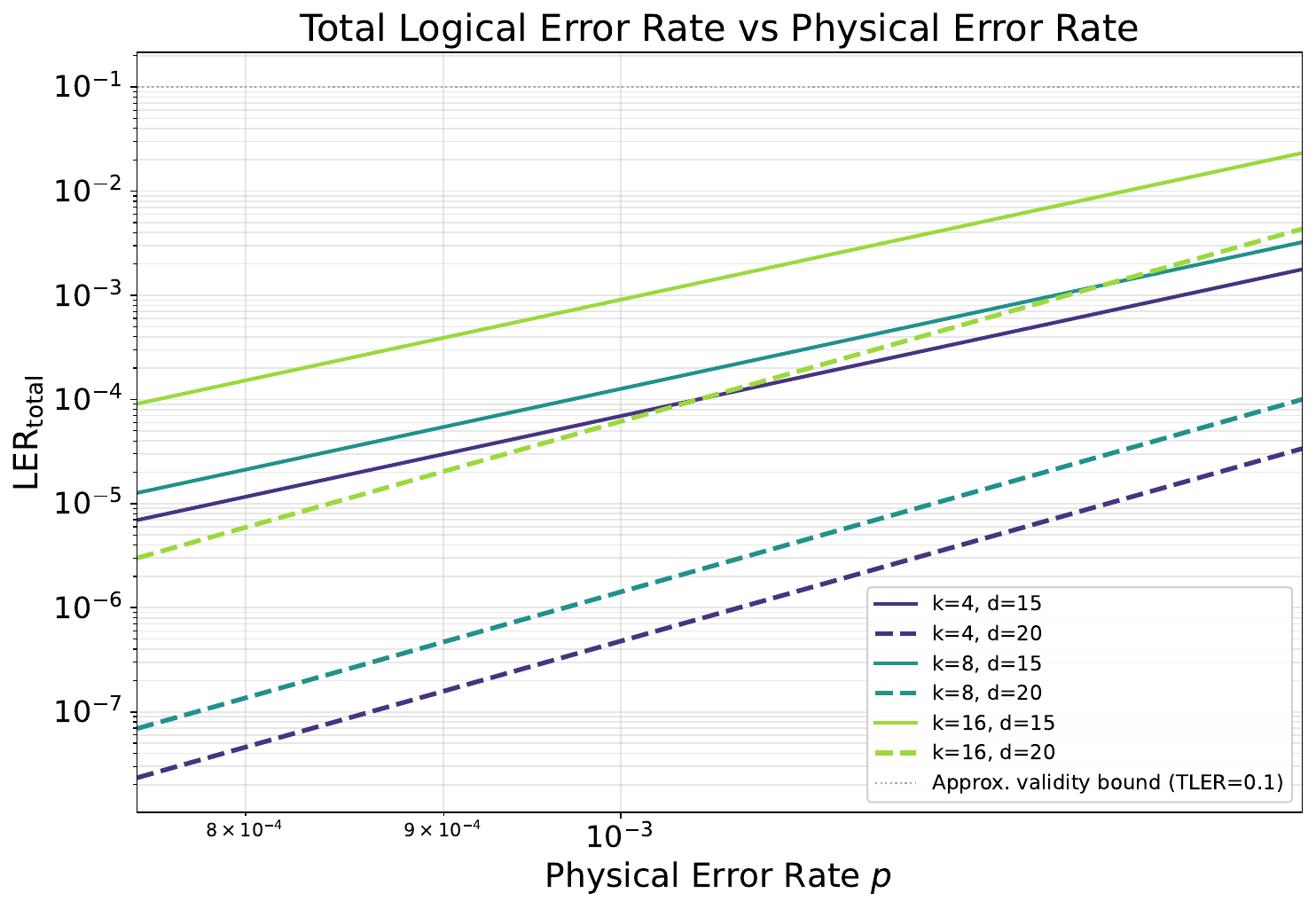}
    \caption{Total logical error rate (TLER) versus physical error rate $p$ for distributed 
surface-code circuits (30 qubits, 200 gates, $n_{\text{ancilla}}=5$, $w=5$, $\lambda=4$) 
at code distances $d \in \{15, 20\}$ and idle noise multipliers 
$k = T_{\text{EPR}}/T_{\text{syn}} \in \{4, 8, 16\}$. 
The dashed horizontal line marks $\text{TLER}=0.1$, above which the 
additive assumption in (Eq.~\ref{eq:tler_full}) loses validity.}
    \label{fig:tradeoff_k}
\end{figure}
 We evaluate Eq.~\ref{eq:tler_full} across a range of physical error rates, code distances and idle noise multipliers $k = T_{\text{EPR}}/T_{\text{syn}}$ to examine how the interplay between code distance and network entanglement generation rate shapes the logical error. We have fixed $\lambda$ which is related to the extra noise induced by the Bell pairs at the seam, see Appendix~\ref{appendix:noise_in_lattice_surgery}. 

For a fixed $k$ (equivalently, a fixed entanglement generation rate $1/T_{\text{EPR}}$), increasing the code distance consistently reduces the total logical error rate across the plotted range. This is the expected sub-threshold behavior: at any fixed noise profile, larger $d$ provides stronger error suppression. However, in a distributed architecture, $k$ is not independent of $d$. Each syndrome extraction round of a remote lattice-surgery operation requires $O(d)$ Bell pairs to be delivered by the network. Considering the limited Bell pair generation rate within the communication module, the remote entanglement distribution rate between two QPUs is capped. Therefore, the amount of time to serve $d$ Bell pairs for each syndrome round is increasing as a function of $d$.\par 

This coupling between $d$ and $k$ changes how we approach distance selection. In a monolithic system, all operations are local, the noise profile is independent of $d$, and increasing $d$ is beneficial below threshold. In a distributed system, increasing $d$ simultaneously improves per-gate error suppression and increases the network load, potentially pushing the system into a higher-$k$ regime where the idle noise accumulates rapidly. Figure~\ref{fig:tradeoff_k} illustrates the consequence of this coupling. We do observe that for fixed $k$, increasing $d$ is helpful. But the interesting point to observe here is the cross-over. Reading across the curves in the figure, the $d = 20$ curve at $k = 16$ crosses above both the $d = 15$ curves at $k = 4$ and $k = 8$. While the location of the cross-over is dependent on the modeling assumptions, the takeaway here is the \textit{existence} of the crossover. In this regime, higher code distance implies higher Bell pair need, and this results in a network-induced penalty which offsets the improved error suppression offered by increasing the distance.

%% file: appendix_section_compilation.tex
\appendix

\section{Compilation and Scheduling Algorithms}
\label{appendix:algorithms}

This appendix provides the pseudocode for the compilation and scheduling 
procedures summarized in Section~\ref{sec:latency_model}.

\subsection{All-to-All Compilation}
\label{appendix:all_to_all_compilation}

Algorithm~\ref{alg:l2p} formalizes the iterative ancilla assignment and 
lattice-surgery expansion procedure described in 
Section~\ref{subsec:logical_to_physical}. For each layer of the logical 
circuit DAG, CNOT gates are assigned ancilla patches from the free pool 
through repeated passes with ancilla reuse, then expanded into annotated 
$M_{ZZ}$ and $M_{XX}$ parity measurements. In Algorithm~\ref{alg:l2p}, $\mathsf{FindShell}(G, p)$ returns the set of ancilla patches reachable from patch $p$ in the connectivity graph $G$. 

\begin{algorithm}[H]
\caption{Ancilla assignment and lattice-surgery expansion (all-to-all connectivity)}
\label{alg:l2p}
\begin{algorithmic}[1]
\REQUIRE $DAG$: logical circuit;
         $G$: connectivity graph;
         $M$: logical $\rightarrow$ physical patch mapping;
         $A$: ancilla patches;
         $\mathsf{FindShell}(G, p)$: reachable ancillas from patch $p$.
\ENSURE $DAG_{\text{phys}}$

\FOR{each layer $L$ in $DAG$}

  \STATE Copy single-qubit gates in $L$ to $DAG_{\text{phys}}$
  \STATE $U \gets$ CNOT gates in $L$
  \STATE $\text{Assigned} \gets \emptyset$

  \COMMENT{Iterative ancilla assignment with reuse}
  \REPEAT
    \STATE Reset all ancillas in $A$ to \textsc{Free}
    \STATE $U' \gets \emptyset$
    \FOR{each $cx(q_c, q_t) \in U$}
      \STATE $S \gets \mathsf{FindShell}(G, M(q_c)) \cap 
             \mathsf{FindShell}(G, M(q_t))$
      \IF{$\exists$ free $a \in S$}
        \STATE $\text{Assigned} \gets \text{Assigned} \cup \{(cx, a)\}$
        \STATE Mark $a$ as \textsc{Busy}
      \ELSE
        \STATE $U' \gets U' \cup \{cx\}$
      \ENDIF
    \ENDFOR
    \STATE $U \gets U'$
  \UNTIL{$U = \emptyset$}

  \COMMENT{Expand all assigned CNOTs}
  \FOR{each $(cx(q_c, q_t),\; a) \in \text{Assigned}$}
    \STATE Emit $M_{ZZ}(q_c, a)$ and $M_{XX}(q_t, a)$, each 
           labeled \textbf{local} or \textbf{remote} by QPU location
    \STATE Emit ancilla measure/reset for $a$
  \ENDFOR

\ENDFOR

\RETURN $DAG_{\text{phys}}$
\end{algorithmic}
\end{algorithm}

\subsection{Network-Aware Scheduler}
\label{appendix:scheduler}

Algorithm~\ref{alg:ftdqc_scheduler} formalizes the event-driven scheduling 
procedure described in Section~\ref{subsec:scheduling}. The scheduler 
maintains per-qubit availability tracking and models network resource 
contention for remote operations.

\begin{algorithm}[H]
\caption{Event-driven network-aware scheduler}
\label{alg:ftdqc_scheduler}
\begin{algorithmic}[1]
\REQUIRE Annotated DAG $DAG_{\text{phys}}$, network graph $G$, 
         timing model $T$
\ENSURE Circuit makespan $t_{\text{total}}$

\STATE $t \gets 0$, initialize priority queue $Q$
\STATE Initialize qubit availability tracker: 
       $t_{\text{free}}[q] \gets 0$ for all qubits $q$

\WHILE{$DAG_{\text{phys}}$ has unfinished gates}

\STATE  \COMMENT{Schedule all ready operations}
  \FOR{each gate $op$ in the frontier of $DAG_{\text{phys}}$}
    \STATE $t_{\text{start}} \gets \max(t,\; 
           \max_{q \in op} t_{\text{free}}[q])$

    \IF{$op$ is local}
      \STATE $t_{\text{end}} \gets t_{\text{start}} + 
             T_{\text{local}}(op)$
    \ELSE
      \IF{required network resources are available}
        \STATE Reserve path
        \STATE $t_{\text{end}} \gets t_{\text{start}} + 
       2d \cdot T_{\text{syn}} + 2d \cdot \max(T_{\text{syn}}, T_{\text{EPR}})$
      \ELSE
        \STATE Skip $op$; retry at next scheduling step
        \STATE \textbf{continue}
      \ENDIF
    \ENDIF

    \STATE $t_{\text{free}}[q] \gets t_{\text{end}}$ for all 
           qubits $q \in op$
    \STATE $Q.\text{push}(op,\; t_{\text{end}})$
  \ENDFOR

  \COMMENT{Advance time to next completing operation}
  \STATE $t \gets$ earliest completion time in $Q$
  \STATE Pop all operations completing at $t$; remove from 
         $DAG_{\text{phys}}$
  \STATE Release network resources held by completed operations

\ENDWHILE

\RETURN $t$
\end{algorithmic}
\end{algorithm}

\subsection{Grid-Based Compilation}
\label{appendix:grid_compilation}

In the grid-based architecture, logical patches are laid out in a 
checkerboard pattern of data and ancilla patches. Implementing a logical 
CNOT now requires not only choosing an ancilla with the correct boundary 
type, but also moving the data qubits via SWAPs to grid locations where 
the control and target can both interact with the appropriate ancilla or 
communication module through compatible boundaries.

We distinguish two cases:
(i)~\emph{intra-module} routing, where both qubits reside within the same 
module, and
(ii)~\emph{inter-module} routing, where the control and target lie on 
different modules and must use dedicated communication modules.

\paragraph{Intra-module routing.}

Given a CNOT gate between qubits $q_1$ and $q_2$ 
residing in the same module, we construct a SWAP-adjacency graph
$G_{\text{swap}}$ whose vertices are locations for the surface code 
patches and where an edge $(u,v)$ exists if there is an ancilla $a$ that 
can mediate a SWAP between $u$ and $v$ (i.e., $u$ and $v$ couple to $a$ 
with complementary boundary types). We then enumerate all \emph{CX-ready 
configurations} for the fixed control location $p(q_1)$: pairs $(v, a)$ 
such that $p(q_1)$ and $v$ are both adjacent to ancilla $a$ with 
complementary boundary types, enabling a lattice-surgery CNOT. A 
shortest-path search in $G_{\text{swap}}$ then determines the minimum 
number of SWAPs required to move $q_2$ from its current location $p(q_2)$ 
to any CX-ready location $v$. Algorithm~\ref{alg:intra_module} details 
this procedure.

The SWAP path is expanded into a sequence of SWAP operations and inserted 
into the DAG, followed by the CNOT expansion into $M_{ZZ}$ and $M_{XX}$ 
measurements using the chosen ancilla. Each SWAP decomposes into three 
sequential lattice-surgery CNOTs. In the checkerboard layout, the complementary boundary requirement 
forces any valid data--ancilla--data configuration into an L-shaped 
arrangement, ensuring that the two data patches that share an ancilla 
$a$ necessarily also share a second ancilla $a'$ whose boundary 
orientations are exactly complementary to those of $a$. The outer two CNOTs use $a$ and the middle CNOT (which reverses 
control and target) uses $a'$, so no intermediate Hadamard operations are 
required. Each measurement in the intra-module sequence is classified as 
local, since both qubits reside within the same module.

\begin{algorithm}[H]
\caption{Intra-module CX routing on a grid}
\label{alg:intra_module}
\begin{algorithmic}[1]
\REQUIRE Grid graph $G$, fixed qubit location $p(q_1)$, 
         movable qubit location $p(q_2)$, ancilla set $A$
\ENSURE SWAP path moving $q_2$ to a CX-ready location w.r.t.\ $q_1$,
        and the ancilla $a^*$ for the final CX

\STATE $G_{\text{mod}} \gets$ subgraph of $G$ restricted to the 
       module containing $p(q_1)$

\COMMENT{Step 1: build SWAP-adjacency graph}
\STATE Initialize $G_{\text{swap}}$ with data locations in 
       $G_{\text{mod}}$ as vertices
\FOR{each pair of data locations $(u, v)$ in $G_{\text{mod}}$}
  \IF{$\exists\, a \in A$ adjacent to both $u$ and $v$ in 
      $G_{\text{mod}}$, with complementary boundary types}
    \STATE Add edge $(u, v)$ to $G_{\text{swap}}$ 
           with mediator $a$
  \ENDIF
\ENDFOR

\COMMENT{Step 2: enumerate CX-ready configurations for $q_1$}
\STATE $C \gets \emptyset$
\FOR{each ancilla $a$ adjacent to $p(q_1)$ in $G_{\text{mod}}$}
  \STATE $\tau \gets$ boundary type of edge $(p(q_1), a)$
  \IF{$\tau$ is a valid lattice-surgery boundary}
    \FOR{each data location $v \neq p(q_1)$ adjacent to $a$}
      \IF{boundary type of $(v, a)$ is complementary to $\tau$}
        \STATE $C \gets C \cup \{(v, a)\}$
      \ENDIF
    \ENDFOR
  \ENDIF
\ENDFOR

\COMMENT{Step 3: shortest SWAP path to any CX-ready location}
\STATE $(v^*, a^*) \gets \arg\min_{(v,a) \in C}\; 
       \text{dist}(G_{\text{swap}},\; p(q_2),\; v)$
\STATE $\text{Path} \gets \text{ShortestPath}(G_{\text{swap}},\; 
       p(q_2),\; v^*)$

\RETURN Path with mediating ancillas, and CX ancilla $a^*$
\end{algorithmic}
\end{algorithm}

\paragraph{Inter-module routing.}
When the control $q_1$ and target $q_2$ reside on different modules, we must 
additionally route through communication modules and the inter-module 
network. By convention, the ancilla patch is always co-located with the 
control qubit $q_1$ on module $A$. This fixes the local measurement as 
$M_{ZZ}(q_1, a_A)$ and the remote measurement as $M_{XX}(q_2, a_A)$, 
mediated by communication modules $c_A$ on module $A$ and $c_B$ on 
module $B$. Since $q_1$ and $q_2$ may not initially be adjacent to the 
required ancilla or communication modules, they must first be routed via 
SWAPs to suitable target locations $q_1^t$ and $q_2^t$ respectively. The boundary type requirements follow from the lattice-surgery 
protocol: $q_1^t$ must couple to $a_A$ with a Z-type boundary, while 
$a_A$ must couple to $c_A$ and $q_2^t$ must couple to $c_B$ with X-type 
boundaries.

The inter-module routing problem is to find a \emph{communication chain}
\[
q_1 \;\rightarrow\; q_1^{t} \;\rightarrow\; a_A \;\rightarrow\; c_A 
\;\leadsto\; c_B \;\rightarrow\; q_2^{t}
\]
such that:
(i)~$q_1$ can be moved (via intra-module SWAPs) to a location $q_1^{t}$ 
that couples to a local ancilla $a_A$ with a Z-type boundary, where $a_A$ 
is in turn adjacent to a communication module $c_A$ with an X-type 
boundary;
(ii)~$c_A$ and $c_B$ are connected through the inter-module network; and
(iii)~$q_2$ can be moved to a location $q_2^{t}$ that couples to $c_B$ 
with an X-type boundary.

\noindent
Condition~(ii) is guaranteed by the Fat-tree topology 
(Section~\ref{subsec:fat_tree}), which provides connectivity between any 
pair of communication modules. 
The search therefore focuses on satisfying conditions~(i) and~(iii) while 
minimizing routing overhead.

We reuse the intra-module SWAP-adjacency graphs $G_{\text{swap}}^A$ and 
$G_{\text{swap}}^B$ constructed by Algorithm~\ref{alg:intra_module} for 
each module. The search enumerates candidate chains over all valid 
combinations of ancilla and communication modules on module $A$, 
communication modules on module $B$, and target locations, minimizing the 
total SWAP cost $d_1 + d_2$, where $d_1$ and $d_2$ are the shortest-path 
distances needed to move $q_1$ and $q_2$ to their respective targets in 
$G_{\text{swap}}^A$ and $G_{\text{swap}}^B$. 
Algorithm~\ref{alg:inter_module} details this procedure.
\begin{figure}[htp!]
    \centering
    \includegraphics[width=0.99\linewidth]{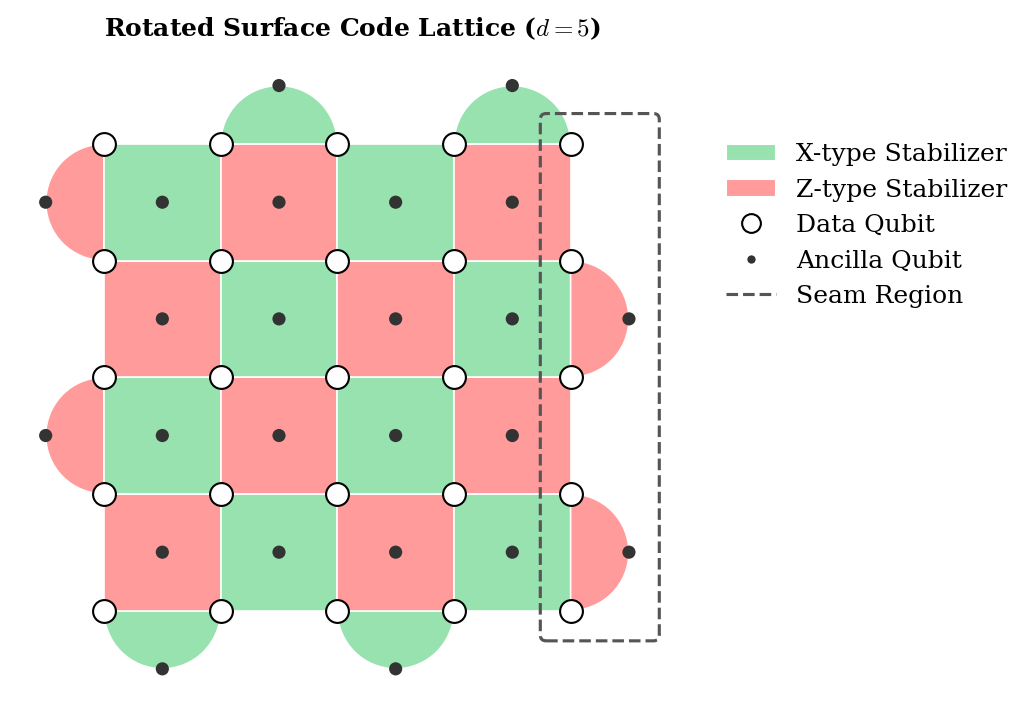}
    \caption{Rotated surface-code having distance ($d=5$) with the seam region indicated by a dashed rectangle. 
Green and red plaquettes denote $X$-type and $Z$-type stabilizers, respectively. 
Open circles represent data qubits, while filled dots represent ancilla qubits. 
The dashed rectangle marks the seam region along the right edge, where the corresponding data and ancilla qubits are assumed to experience elevated noise.}
    \label{fig:seam_visualization}
\end{figure}
For the grid architecture, the compilation step is therefore more involved 
than in the all-to-all case: we must respect boundary types, choose 
appropriate ancillas and communication modules, and insert the SWAP 
sequences as required. The resulting modified DAG includes both the SWAP 
operations and the local/remote parity measurements. Once this routing has 
been applied, the subsequent scheduling stage treats grid-based and 
all-to-all architectures uniformly through their induced sets of local and 
remote operations.

\begin{algorithm}[H]
\caption{Inter-module CX routing via communication modules (grid)}
\label{alg:inter_module}
\begin{algorithmic}[1]
\REQUIRE Global graph $G$, initial locations $p(q_1)$ on module $A$ 
         and $p(q_2)$ on module $B$, communication modules $C$, 
         ancilla set $A$
\ENSURE Optimal target locations $q_1^t$, $q_2^t$, ancilla $a_A$, 
        communication modules $c_A$, $c_B$, and total SWAP cost 

\STATE Build $G_{\text{swap}}^A$, $G_{\text{swap}}^B$ for each module 
       (as in Algorithm~\ref{alg:intra_module})
\STATE $\text{Cost}_{\min} \gets \infty$
\STATE $\tau \gets ZZ$;\; $\tau' \gets$ complementary boundary type of $\tau$

\FOR{each ancilla $a_A$ in module $A$}
  \FOR{each data location $q_1^t$ adjacent to $a_A$ with type $\tau$}
    \STATE $d_1 \gets \text{ShortestPath}(G_{\text{swap}}^A,\; 
           p(q_1),\; q_1^t)$
    \IF{no path} \STATE\textbf{continue} \ENDIF
    \FOR{each comm.\ module $c_A$ adjacent to $a_A$ with type $\tau'$}
      \FOR{each comm.\ module $c_B$ in module $B$}
        \FOR{each data location $q_2^t$ adjacent to $c_B$ with 
             type $\tau'$}
          \STATE $d_2 \gets \text{ShortestPath}(G_{\text{swap}}^B,\; 
                 p(q_2),\; q_2^t)$
          \IF{no path} \STATE \textbf{continue} \ENDIF
          \IF{$d_1 + d_2 < \text{Cost}_{\min}$}
            \STATE Update $\text{Cost}_{\min}$ and record 
                   $(q_1^t, q_2^t, a_A, c_A, c_B)$
          \ENDIF
        \ENDFOR
      \ENDFOR
    \ENDFOR
  \ENDFOR
\ENDFOR

\RETURN best configuration, $\text{Cost}_{\min}$
\end{algorithmic}
\end{algorithm}

%% file: appendix_custom.tex
\section{ Asymmetric Circuit-Level Noise Characterization}
\label{appendix:noise_in_lattice_surgery}

Surface-code simulations typically employ a circuit-level noise model in which independent stochastic Pauli errors are injected into every physical operation, including quantum gates, state initialization, and measurement. The standard model is parameterized by a single physical error rate $p$, applied homogeneously as follows:

\begin{enumerate}
    \item \textbf{Gate noise:} A depolarizing Pauli error with probability $(p)$ after every single-qubit and two-qubit Clifford gate, spread uniformly over the corresponding non-identity Pauli errors.
    \item \textbf{Initialization noise:} A basis-dependent Pauli error is applied with probability ($p$) after each qubit reset.
    \item \textbf{Measurement noise:} A basis-dependent Pauli error with probability ($p$) preceding each measurement.
    \item \textbf{Idle decoherence:} A single-qubit depolarizing channel with probability $(p)$ is applied to each data qubit prior to every syndrome-extraction round.
\end{enumerate}

This homogeneous model provides a useful baseline but does not capture two features intrinsic to distributed architectures. First, remote entanglement generation introduces latency that extends idle periods during syndrome extraction, increasing decoherence beyond what the uniform rate $p$ accounts for. Second, qubits at module boundaries rely on entanglement links whose fidelity is generally lower than that of local operations. Prior work has addressed these effects partially: Ramette et al.~\cite{Ramette2024} model elevated noise on seam qubits at inter-module boundaries, and Jacinto et al.~\cite{Jacinto2026} propose syndrome-extraction strategies that preserve code distance in distributed settings. However, neither accounts for the additional decoherence arising from entanglement generation wait times. To capture both effects jointly, we work with the following modified noise model.

\paragraph{Model definition.}
All standard circuit-level channels retain the base rate $p$, with two modifications:
\begin{enumerate}
    \item \textbf{Idle noise rescaling:} Data-qubit idle depolarization is elevated to $p_I = k\, p$ (with $k \geq 1$) to account for the extended idle periods caused by entanglement generation latency.
    \item \textbf{Seam noise injection:} An additional single-qubit depolarizing channel with rate $p_s = \lambda\, p$ (with $\lambda \geq 1$) is added  once per syndrome extraction round to both data and ancilla qubits within the designated seam region, see Figure~\ref{fig:seam_visualization}, capturing the reduced fidelity of inter-module entanglement links.
\end{enumerate}
The two parameters $k$ and $\lambda$ are independent, enabling systematic exploration of the interplay between latency-induced and interconnect-induced noise. We use \texttt{Stim} for circuit generation and \texttt{PyMatching} for minimum-weight perfect matching (MWPM) decoding.
We assign additional noise to seam qubits to capture the effect of entanglement mediated operations between quantum processors. In practice, the exact noise contribution depends on the syndrome extraction procedure and on the qubit placement across processors. Since the resulting noise is implementation dependent, we consider a simplified noise model for the seam qubits.

\begin{table}[t]
\centering
\caption{Sub-threshold scaling parameters for selected $(k, \lambda)$ configurations.}
\label{tab:scaling_params}
\begin{tabular}{cc|ccccc}
\hline\hline
$k$ & $\lambda$ & $\alpha$ & $\sigma_\alpha$ & $R^2$ & $p_{\mathrm{th}}$ (\%) & $A$ \\
\hline
2  & 1 & 0.526 & 0.004 & 0.997 & 0.665 & 0.054 \\
2  & 4 & 0.530 & 0.005 & 0.995 & 0.665 & 0.058 \\
2  & 8 & 0.537 & 0.008 & 0.986 & 0.664 & 0.063 \\
\hline
4  & 1 & 0.516 & 0.004 & 0.996 & 0.541 & 0.066 \\
4  & 4 & 0.517 & 0.004 & 0.996 & 0.539 & 0.067 \\
4  & 8 & 0.527 & 0.006 & 0.992 & 0.537 & 0.073 \\
\hline
16 & 1 & 0.494 & 0.005 & 0.994 & 0.272 & 0.099 \\
16 & 4 & 0.491 & 0.004 & 0.996 & 0.272 & 0.099 \\
16 & 8 & 0.491 & 0.004 & 0.995 & 0.272 & 0.100 \\
\hline\hline
\end{tabular}
\end{table}

\paragraph{Threshold degradation under distributed noise.}

\begin{figure}[H]
    \centering
    \makebox[\textwidth][c]{\includegraphics[width=0.8\textwidth]{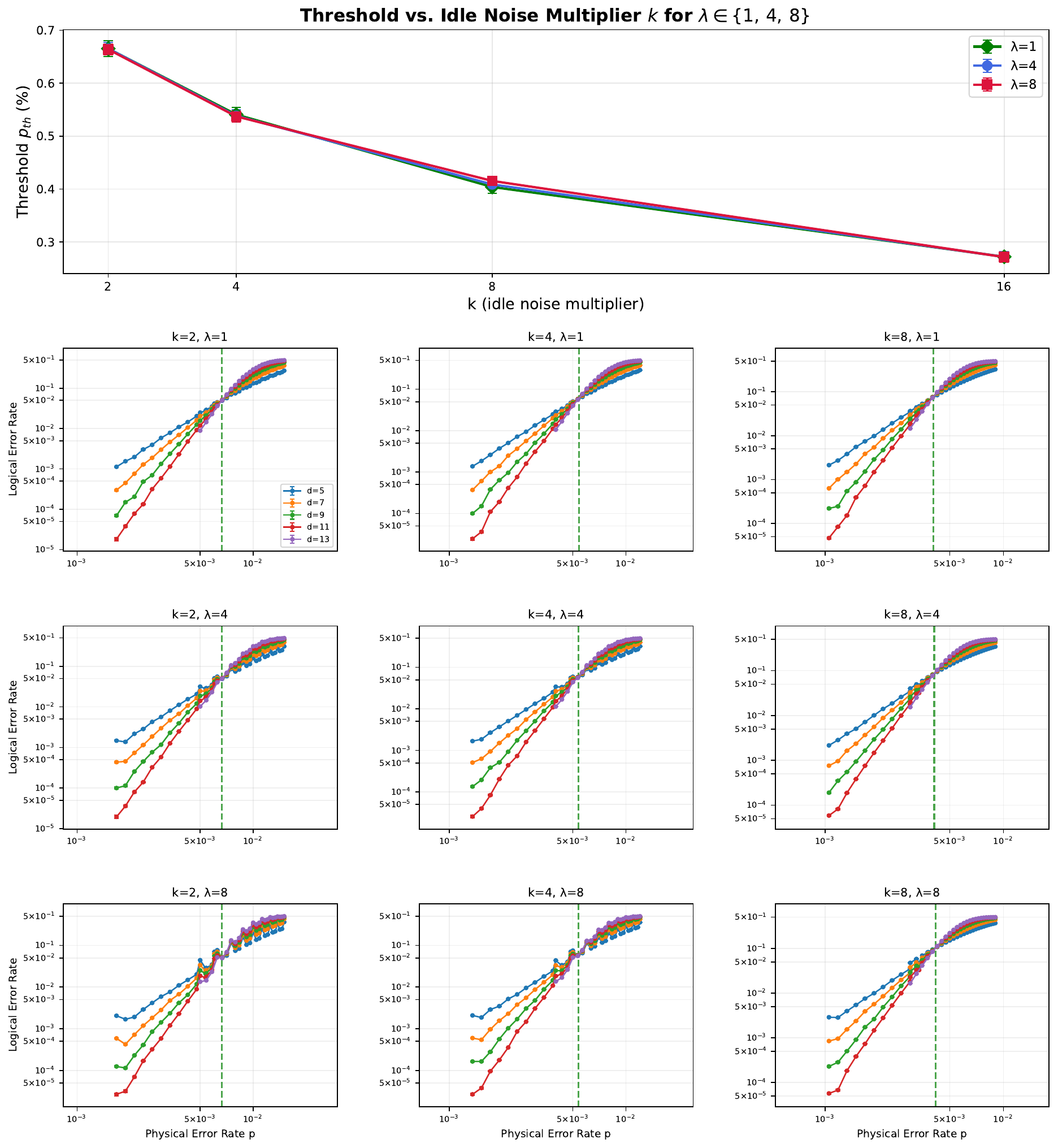}}
    \makebox[\textwidth][c]{%
\begin{minipage}{0.8\textwidth}
    \caption{Fault-tolerance threshold and sub-threshold scaling under 
distributed noise. Threshold $p_{\mathrm{th}}$ 
versus idle noise multiplier $k$ for seam noise parameters 
$\lambda \in \{1, 4, 8\}$. Increasing $k$ monotonically reduces the 
threshold across all $\lambda$ values. 
For the remaining panels, we show the logical error rate $p_L$ versus physical 
error rate $p$ for code distances $d \in \{5, 7, 9, 11, 13\}$ under 
selected $(k, \lambda)$ configurations. Crossing points indicate the 
fault-tolerance threshold for each configuration.}
\label{fig:threshold_sweep}

\end{minipage}}
\end{figure}

Figure~\ref{fig:threshold_sweep} quantifies the fault-tolerance threshold as a function
of the idle noise multiplier $k$ and 
$\lambda \in \{1, 4, 8\}$. For any fixed $\lambda$,
increasing $k$ monotonically reduces the threshold, confirming that entanglement
generation latency which stretches idle periods during syndrome extraction, directly
erodes the tolerance.

To each $k$ and $\lambda$, we fit the logical error rate to the standard scaling model $p_L = A \cdot (p/p_{th})^{\alpha d}$, where $p_{th}$ is extracted via the crossing-point method across adjacent distance pairs. From the fitted parameters, we have $A$, $\alpha$, and $p_{th}$ for each $(k, \lambda)$. So we can extrapolate the logical error rate for any $(k, \lambda, d, p)$ using:
 
$$p_L = A \cdot \left(\frac{p}{p_{th}}\right)^{\alpha \cdot d}$$.

\onecolumngrid

\twocolumngrid